\documentclass[prd,twocolumn,superscriptaddress,floatfix,amsmath,amssymb]{revtex4-1}

\usepackage{float} 

\usepackage[normalem]{ulem}
\usepackage[english]{babel}
\usepackage{graphicx}
\usepackage{dcolumn}
\usepackage{bm}
\usepackage{verbatim}
\usepackage{mathrsfs}
\usepackage{amsmath}
\usepackage{cancel}
\usepackage{epstopdf}
\usepackage{mathtools}
\usepackage{color}
\newcommand{\proj}[2]{\left| {#1} \right\rangle\!\left\langle {#2} \right|}
\newcommand{\ket}[1]{\left| {#1} \right\rangle}
\newcommand{\bra}[1]{\left\langle {#1} \right|}

\newcommand{\dd}{\mathrm{d}}
\newcommand{\ii}{\mathrm{i}}
\newcommand{\II}{\textsc{i}}

\newcommand{\abs}[1]{\vert #1\vert}

\usepackage[usenames,dvipsnames]{pstricks}
\usepackage{epsfig}
\usepackage{pst-grad} 
\usepackage{pst-plot} 

\usepackage{hyperref}

\usepackage{verbatim}

\begin{document}

\title{Relativistic Quantum Optics:\\ The relativistic invariance of the light-matter interaction models}

\author{Eduardo Mart\'{i}n-Mart\'{i}nez}
\email{emartinmartinez@uwaterloo.ca}
\affiliation{Institute for Quantum Computing, University of Waterloo, Waterloo, Ontario, N2L 3G1, Canada}
\affiliation{Department of Applied Mathematics, University of Waterloo, Waterloo, Ontario, N2L 3G1, Canada}
\affiliation{Perimeter Institute for Theoretical Physics, Waterloo, Ontario, N2L 2Y5, Canada}

\author{Pablo Rodriguez-Lopez}
\affiliation{Materials Science Factory, Instituto de Ciencia de Materiales de Madrid, ICMM-CSIC, Cantoblanco, E-28049 Madrid, Spain}
\affiliation{GISC-Grupo Interdisciplinar de Sistemas Complejos, 28040 Madrid, Spain}

\begin{abstract}
In this note we discuss the invariance under general changes of reference frame of all the physical predictions of particle detector models in quantum field theory in general and, in particular, of those used in quantum optics to model atoms interacting with light. We find explicitly how the light-matter interaction Hamiltonians change under general coordinate transformations, and analyze the subtleties of the Hamiltonians commonly used to describe the light-matter interaction when relativistic motion is taken into account.
\end{abstract}

\maketitle

\section{Introduction}

Particle detector models may be thought of as localized, controllable, first quantized systems that couple locally in space and time to quantum fields. Particle detector models in quantum field theory were pioneered by Unruh and DeWitt \cite{unruh_notes_1976,DeWitts}, and are used in the literature on quantum field theory, as well as representing atoms coupled to the electromagnetic field in  the description of the light-matter interaction in quantum optics \cite{Glauber,JaynesCumm,scullybook}.

On the one hand, from the fundamental field-theoretical point of view, particle detectors simplify the task of extracting localized information about the field without resorting to  projective measurements of  localized field observables \cite{Lin2,Drago1,Drago2,Lin2014773}. Particle detector models have been successfully employed  in a host of contexts in fundamental quantum filed theory \cite{Crispino,Takagi}. Perhaps one of the best-known ones is the operational formulation of the Hawking and Unruh effects (see, e.g., \cite{unruh_notes_1976,candelas_irreversible_1977}).

On the other hand, from the more applied point of view, more or less elaborated or simplified versions of particle detector models are ubiquitous to model the light-matter interaction in experimental setups in quantum optics \cite{scullybook} and in superconducting circuits \cite{Wallraff:2004aa}. For example, an alkali atom as a first quantized system, can serve as such a detector for the second quantized electromagnetic field. In fact, the common light-matter interaction models, such as for instance the Glauber model \cite{Glauber}, or the Jaynes-Cummings model and its variants \cite{JaynesCumm}, are in essence simplifications of the Unruh-DeWitt (UDW) model \cite{DeWitts}, with extra approximations made on them (See, e.g., \cite{Martin-Martinez2015} for details). Indeed, as shown in section 2 of \cite{Pozas-Kerstjens2016}, the UDW model is a good model of dipolar coupling in the light-matter interaction in quantum optics.

Particle detectors (both in the UDW variant and even more elaborate realistic atomic models coupling to the electromagnetic fields, e.g.,  \cite{Pozas-Kerstjens2016,Richard}) have also been extensively used  in  the field of relativistic quantum information and quantum field theory. Examples can be found in relativistic quantum computing \cite{AasenPRL,Lee,Chris}, quantum communication via field quanta \cite{mathieuachim1,Robort2,Landulfo,Robort3}, cosmology \cite{Gibbons1977,cosmoq,FilCos,QuanG,Blasco}, the study of Casimir-Polder interactions \cite{Alhambra2014,Marino2014,Dalvitian}  and in a number of studies of effects related to the presence of spacelike and timelike entanglement in the vacuum state of quantum fields both from fundamental \cite{Valentini1991,Reznik2003,Reznik1,VerSteeg2009,Olson2011,Oslon2,Salton:2014jaa,Pozas-Kerstjens:2015,Yasuada}  and applied \cite{Farming,Brown:2014en} perspectives. Interestingly, in these studies it is shown that it is possible to harvest correlations from the field vacuum to spacelike separated detectors, which gives an operational proof of the spacelike entanglement present in the quantum vacuum \cite{Alegbra1,Alegbra2}.

Because of the fundamental and applied usefulness of particle detector models in the context of relativistic quantum information, the natural question arises as to what extent these models (which involve non-relativistic systems coupled to fully relativistic quantum fields) behave in a covariant way in regimes where relativistic effects become important.  This is of special importance when studying phenomena for which the causal behaviour of the model is key. Although particle detector models have been proven not to suffer from faster-than light signalling \cite{mathieuachim1,Martin-Martinez2015}, showing that is not sufficient: the models would not be any good if all their physical predictions were not be invariant under changes of reference frame.  

The purpose of this note is double. On the one hand we will show (non-perturbatively) how the transformations between reference frames do not modify the predictions of properly formulated particle detector models, which is a manifest consequence of the fact that the coupling between the detector and the field is fully covariant regardless of the first-quantized nature of the detector. This will allow us to discuss the subtleties on the interpretation of the different parts of the Hamiltonian when particle detectors undergo relativistic trajectories. For example, when we are trying to model the physics of moving detectors. 

On the other hand, we will show explicitly, and in a constructive way, how to perform general coordinate transformations on the light-matter interaction Hamiltonians.  This will allow us to constructively show what is the exact form of the Hamiltonians in different reference frames, something that is not shown explicitly in the literature to the authors' knowledge. This is especially important in the case of smeared detectors (which, like atoms, are not pointlike and instead have a spatial profile), where the relationship between Hamiltonians in different reference frames presents subtleties due to the mixing of space and time in general coordinate transformations.

After this, and as a didactic exercise, we will illustrate the invariance explicitly with an example: a perturbative calculation of a transition probability in two reference frames for pointlike and for smeared detectors.

\section{Time reparametrizations}

Before talking about relativity or particle detectors, let us first clarify a more basic basic question: how a Hamiltonian behaves under time reparametrization. That is, given a Hamiltonian $\hat{H}^{t}(t)$ that generates time translations with respect to a time parameter $t$, what is the Hamiltonian $\hat{H}^{\tau}(\tau)$ that generates translations with respect to a different time parameter $\tau$, knowing that the relationship between the two parameters is given by the non-singular reparametrization function $t(\tau)$.

Of course this is well-known (it is just a basic problem of classical mechanics) and can be found elsewhere (e.g. \cite{Brown:2014en} and other places such as analytical mechanics texts), but we include here a very easy way to see the effect of a reparametrization on a Hamiltonian just from the transformation of Schr\"odinger equation. Namely, if we start from the equation
\begin{equation}
   \ii \hbar \frac{\dd}{\dd t}\ket{\psi}=\hat{H}^{t}(t)\ket{\psi}
\end{equation}
and carry out the reparametrization to a new time parameter $\tau$, using chain rule for the time derivative we get
\begin{equation}\label{provisioH}
   \ii \hbar \dfrac{\dd \tau}{\dd t}\dfrac{\dd}{\dd \tau}\ket{\psi}=\hat{H}^{t}(t(\tau))\ket{\psi}.
\end{equation}
Notice that $\hat{H}^{t}(t(\tau))$ is not the Hamiltonian that generates time translations with respect to $\tau$. Rather, it still is the Hamiltonian that generates translations with respect to $t$ written as a function of $\tau$. To obtain $\hat{H}^\tau(\tau)$ we need to rewrite \eqref{provisioH} in the standard form of the Schr\"odinger equation, which yields
\begin{equation}\label{provisioH2}
   \ii \hbar \dfrac{\dd}{\dd \tau}\ket{\psi}= \dfrac{\dd t}{\dd \tau}\hat{H}^{t}(t(\tau))\ket{\psi}.
\end{equation}
This allows us to directly identify the form of the reparametrized Hamiltonian $\hat{H}^\tau(\tau)$ that generates translations with respect to $\tau$:
\begin{equation}\label{ScoringPar}
    \hat{H}^\tau(\tau)=\dfrac{\dd t}{\dd \tau}\hat{H}^{t}(t(\tau)),
\end{equation}
that is, there is a multiplicative factor $\frac{\dd t}{\dd \tau}$ (that we will call from now on a \textit{redshift factor}) resulting from the reparametrization. 

After clarifying this basic question we want to highlight that, as is easy to see,  time evolution is invariant under time reparametrization. Indeed,
\begin{align}\label{reparam}
    \nonumber\hat U &=\mathcal{T}\exp\left[\frac{-\ii}{\hbar}{\int_{\mathbb{R}} \dd \tau\,\hat{H}^\tau(\tau)}\right]\\
    &=\mathcal{T}\exp{\left[\frac{-\ii}{\hbar}\int_{\mathbb{R}} \dd \tau\dfrac{\dd t}{\dd \tau}\hat{H}^{t}(t(\tau))\right]}\nonumber\\
    &=\mathcal{T}\exp{\left[\frac{-\ii}{\hbar}\int_{\mathbb{R}} \dd t\,\hat{H}^{t}(t)\right]}.
\end{align}

Notice that in this note, the region of integration for the time evolution operator is the whole $\mathbb{R}$. The finite nature of the interaction is implemented through the time dependence of the Hamiltonian, and not in the integration limits as is somewhat common in some textbooks. This is more general and also the cleanest way to talk about interactions of detectors with quantum fields, since in this way one can account for the rate of how fast the interactions are turned on and off in the time dependence of the interaction Hamiltonian, instead of suddenly switching it off at a given time. This is particularly relevant in scenarios where sudden switching can induce divergences \cite{Satz2006,Satz2007,Hodgkinsonclick}. 

Recall that in this section we are just considering the reparametrizations, not necessarily coming from general changes of coordinates. Changes of coordinates associated with changes of reference frame requires addressing extra subtleties that will be better clarified when we consider the case of smeared detectors in section \ref{detectoria}.

\section{The detector-field system Hamiltonian}

The Unruh-Dewitt models (and furthermore, all the light-matter interaction models used in quantum optics in general grounds) consist of a particle detector, which is a non-relativistic first-quantized localized system coupled to a quantum field. 

Because of its first-quantized nature, it is not uncommon to hear the following question when it is introduced in relativistic setups: `how is the model going to be respectful with Lorentz covariance if the detector is a non-relativistic first-quantized system?'. In this note, we are going to show that being careful with how the interaction Hamiltonian between field and detector is prescribed, being careful with respect to what time the different parts of the Hamiltonian generate translations, and finally being careful with time reparametrizations associated with changes of reference frame, the predictions of the model are fully independent of the reference frame in which we describe the dynamics of the detector-field system.

We also have the secondary goal of showing explicitly the subtleties that appear when we consider pointlike and non-pointlike smeared detectors (as it is the case with, e.g.,  realistic atoms) interacting for finite times with the field, and we perform transformations of reference frame, showing explicitly how light-matter interaction Hamiltonians are transformed.

\subsection*{Free and Interaction Hamiltonians}

In quantum optics literature it is often the case that Hamiltonians are given as if the free Hamiltonian of detector and field, as well as the interaction Hamiltonian coupling them would generate translations with respect to the same time parameter. This is so because in usual quantum optics there is no concern with respect to what time those Hamiltonians generate translations since there is usually only one reference frame (the lab frame) and the detector does not move or moves non-relativistically.

However, a particle detector can certainly move relativistically with respect to the lab frame, and the light-matter interaction coupling must have the right properties in terms of invariance of physical predictions under changes of reference frame. Therefore we need to be mindful of this when we write the detector-field interaction Hamiltonian.

Let us briefly review the different Hamiltonians involved in the description of the detector-field dynamics.

\begin{itemize}
\item \textbf{Free Hamiltonian of the detector}: This sets the internal energy scale of the detector $\Omega$. That energy scale is the proper energy gap of the detector, and it is prescribed in the comoving reference frame of the detector. Consequently, in its standard form, it generates translations with respect to the detector's proper time $\tau$. In the Schr\"odinger picture it is given by:
\begin{equation}\label{Hdfree}
    _{S}\hat{H}_{d}^{\tau}  =  \hbar\Omega\hat{\sigma}^{+}\hat{\sigma}^{-},
\end{equation}
where $\hat{\sigma}^{\pm}$ are SU(2) ladder operators. 

\item \textbf{Free Hamiltonian of the field}: Picking a particular quantization inertial frame $(t,\bm{x})$, it is most straightforward to write this Hamiltonian as generating time-translations with respect to that frame. Expanding the field in terms of a set of orthonormal modes  (a basis of solutions to the wave equation that we label with the label $\bm k$) the field energy is the sum of the energies of every mode. Subtracting any zero-point energy contribution this is
\begin{equation}\label{Hffree}
    _{S}\hat{H}_{\phi}^{t}  =   \hbar \int \dd {\bm{k}}\,\omega_{\bm{k}}\hat{a}_{\bm{k}}^{\dagger}\hat{a}_{\bm{k}},
\end{equation}
where $\omega_{\bm{k}}=\sqrt{c^{2}\bm  k^{2} +m^{2}c^{4}\hbar^{-2}}$ and $\hat{a}_{\bm{k}},\hat{a}^{\dagger}_{\bm{k}}$ are respectively annihilation and creation operators satisfying canonical delta-commutation rules. For a massless field (as the one we will be focusing on later) $\omega_{\bm{k}}=c\abs{\bm{k}}$.

\item \textbf{Interaction Hamiltonian}: Although there are non-linear variants (see e.g., \cite{Takagi,Takahashi2011,hummer}), the interaction Hamiltonian is usually bilinear in an observable of the field and an observable the detector. The usual prescription in the literature for Unruh-DeWitt detectors (see e.g., \cite{Takagi}) is that the Hamiltonian takes its simplest form when it generates translations with respect to the detector's proper time $\tau$. Specifically, the interaction Hamiltonian for a pointlike UDW detector is usually prescribed by saying that the detector's monopole moment is coupled to the field amplitude along the trajectory of the detector. Namely, (in the Schr\"odinger picture) the UDW interaction Hamiltonian takes the form \cite{Takagi}
\begin{equation}\label{Hi}
    _{S}\hat{H}_{\phi}^{\tau}  = \hbar c\lambda\chi(\tau)\, \hat \mu_\textsc{s}\, \hat\phi_\textsc{s}(\bm{x}(\tau))
\end{equation}
where the monopole moment and the field operator in the Schr\"odinger picture are given respectively as 
\begin{eqnarray}
\hat\mu_{\textsc{s}} & = & \hat{\sigma}^{+} + \hat{\sigma}^{-},\nonumber\\
\hat{\phi}_{\textsc{s}}(\bm{x}) & = & \int\dd^{d}\bm{k}\left[ 
   \hat{a}_{\bm{k}}^{\dagger}u_{\bm{k}}(\bm{x}) + \hat{a}_{\bm{k}}u_{\bm{k}}^{*}(\bm{x})
\right],
\end{eqnarray}
and where $u_{\bm{k}}(\bm{x})$ are the solutions to the spatial part of the wave equation in an arbitrary basis (e.g., if we choose the plane wave basis, $u_{\bm{k}}(\bm{x})=\frac{e^{-\ii \bm{k}\cdot \bm{x}}}{\sqrt{(2\pi)^n 2\omega_{\bm{k}}c^{-1}}}$, where the normalization factor of the solutions to the full wave equation have been absorbed in the spatial part). $\chi(\tau)$ is a switching function, and the coupling constant $\lambda$ has dimensions of $[\text{length}]^{(d-3)/2}$, where we recall $d$ is the number of spatial dimensions.

The monopole coupling to a scalar field amplitude makes the Unruh-DeWitt model a very good approximation for the atom dipole coupling to the electromagnetic field. Not only that, the UDW model is in fact the Hamiltonian used in quantum optics \cite{scullybook}, usually under additional assumptions such as the rotating wave approximation. How the UDW model captures all the fundamental features of the light matter interaction is discussed in detail section II of \cite{Pozas-Kerstjens2016}. Understood as such approximation, we do not have freedom to choose the interaction Hamiltonian, but instead it is the physics of atoms coupled to the electromagnetic field that dictates it. Hence, it is from the atomic centre of mass reference frame that the atom presents a dipole that couples to the electromagnetic field as seen from the centre of mass of the atom. It is in this simple form of a product of a dipole moment coupled to the electric field that UDW Hamiltonian becomes the scalar version of the atomic coupling to the electromagnetic field, and as such the traditional literature prescription of the interaction Hamiltonian \eqref{Hi} is also  natural  to model the light-matter interaction.
\end{itemize}

The unitary transformation from the Schr\"odinger to the interaction picture is \cite{BallentineB}
\begin{equation}
     _{D}\hat{O}= \hat{U}_{0}\, {_{S}\hat{O}}\, \hat{U}_{0}^{\dagger},
\end{equation}
where $_{D}\hat{O}$ is an operator representing an observable $O$ in the interaction picture, ${_{S}\hat{O}}$ is the same operator in the Schr\"odinger picture, and 
\begin{equation}
    \hat{U}_{0}=\mathcal{T}\exp\left[{\frac{\ii}{\hbar}\int \dd \tau\, {_{S}\hat{H}_0^\tau}}\right].
\end{equation}
 $\hat{H}_0^\tau$ is the free Hamiltonian of the system, $\tau$ is a time parameter with respect to which it  generates translations and the integral is an indefinite integral (the integration constant adds an irrelevant phase). Notice that the operator that switches between the two pictures is invariant under time reparametrizations in the same way as the time evolution operator \eqref{reparam}, and indeed
\begin{align}
    \hat{U}_{0}&=\mathcal{T}\exp\left[{\frac{\ii}{\hbar}\int \dd \tau\, {_{S}\hat{H}_0^\tau}}\right]=\mathcal{T}\exp\left[\frac{\ii}{\hbar}\int \dd \tau\dfrac{\dd t}{\dd \tau}\hat{H}_0^t\right] \nonumber\\&=\mathcal{T}\exp\left[\frac{\ii}{\hbar}{\int \dd t\,\hat{H}_0^{t}}\right].
\end{align}
This should not be surprising since the unitary that switches form Schr\"odinger to interaction pictures corresponds to the evolution operator associated with free evolution. Substituting in this expression the free Hamiltonians \eqref{Hdfree} and \eqref{Hffree}, we can find, respectively, the monopole moment operator and the field operator in their respective interaction pictures,
and using the identities
\begin{eqnarray}\label{usefulid}
e^{\ii\theta\hat{\sigma}^{+}\hat{\sigma}^{-}}\hat{\sigma}^{+}e^{-\ii\theta\hat{\sigma}^{+}\hat{\sigma}^{-}} & = & \hat{\sigma}^{+}e^{\ii\theta},\nonumber\\
e^{\ii\theta\hat{a}_{\bm{k}}\hat{a}^{\dagger}_{\bm{k}}}\,\hat{a}^{\dagger}_{\bm{k}}\,e^{-\ii\theta\hat{a}_{\bm{k}}\hat{a}^{\dagger}_{\bm{k}}} & = & \hat{a}^{\dagger}_{\bm{k}}e^{\ii\theta},
\end{eqnarray}
and its Hermitian conjugates we can finally write the interaction Hamiltonian in the interaction picture. The monopole moment of the detector in the interaction picture can be readily found to be
\begin{align}
\hat\mu(\tau) &=e^{\ii\int \dd\tau {_{S}\hat{H}_{d}^{\tau}}  }\hat\mu_{\textsc{s}}e^{-\ii\int \dd\tau {_{S}\hat{H}_{d}^{\tau}}  }  =  e^{\ii\Omega \tau}\hat{\sigma}^{+} + e^{-\ii\Omega \tau}\hat{\sigma}^{-},\nonumber\\
\hat{\phi}(t,\bm{x}) & = e^{\ii\int \dd t {_{S}\hat{H}_\phi^{t}}  }\hat{\phi}_{\textsc{s}}(\bm{x})e^{-\ii\int \dd t {_{S}\hat{H}_\phi^{t}}  }\\&= \int\dd^{d}\bm{k}\left[ 
   \hat{a}_{\bm{k}}^{\dagger}u_{\bm{k}}(\bm{x})e^{\ii\omega_{\bm k} t} + \hat{a}_{\bm{k}}u_{\bm{k}}^{*}(\bm{x})e^{-\ii\omega_{\bm k} t}
\right],\nonumber 
\end{align}
where we have not written the picture subindex for the interaction picture operators to alleviate notation.

\section{Smeared detectors under general coordinate transformations}\label{detectoria}

Modelling non-pointlike detectors (such as atoms) presents some subtleties that have to be addressed carefully. Same as for the pointlike case, we need to decide what is a reasonable interaction Hamiltonian between a detector and the field. We have already discussed that the interaction Hamiltonian would take its simplest form when it is prescribed from the detector's frame. Hence, finding inspiration in atomic physics, as an atom undergoes motion, we can assume that the electromagnetic interaction keeps every point of the atom accelerating with different accelerations so as to keep rigidity. Note that this is an approximation, not valid for extremely high accelerations, but it is a good approximation still for reasonable high accelerations (even larger than $10^{17}g$ \cite{ruso}). Consider a detector with a fixed shape as seen from its centre of mass reference frame $(\tau,\bm \xi)$, undergoing arbitrary motion with respect to the lab frame ($t,\bm x$), which is also the quantization frame for the field. The detector couples to the field in all the points of its smearing simultaneously (in the centre of mass frame). The free Hamiltonians of both field and detector remain the same (although one has to be careful with the different times that they generate translations with respect to) and therefore the transition from Schr\"odinger to interaction picture is analogous to the previous section. 

 What makes the smeared case more challenging than the pointlike case, is that every point of the detector undergoes a different accelerated trajectory to keep rigidity. In other words, the detector moves keeping the same shape in the centre of mass reference frame (Fermi-Walker rigidity) and interacts with a scalar quantum field in all the points of its trajectory. However, under all these considerations, it is not difficult to prescribe the form of the Hamiltonian in the detector's centre of mass frame (precisely because in that frame the detector does not move). The Hamiltonian basically consists of the pointlike Hamiltonian integrated over the `density' function of the detector, called the \textit{smearing function} in the literature, $f(\bm \xi)$. Writing all this into a mathematical expression we get that the interaction Hamiltonian in the interaction picture is 
\begin{equation}\label{sdet0}
    _{D}\hat{H}_{I}^{\tau}= \hbar c\lambda \chi(\tau)\int\dd^n\bm \xi\, f(\bm \xi)\hat\mu(\tau)\hat \phi[t(\tau,\bm \xi),\bm x(\tau,\bm\xi)],
\end{equation}
which is the form used in the literature of particle detectors \cite{Schlicht2004,Satz2006}. Again, it is possible to see that this smeared version of the UDW model is a very faithful approximation to the realistic Hamiltonian that one gets out of the dipole coupling of a hydrogen atom to the electromagnetic field. The smearing function in the case of hydrogenoid atoms is proportional to  the product of the atomic wave functions of the ground and the excited states (see section II of \cite{Pozas-Kerstjens2016} for details and a derivation from first principles).

This Hamiltonian generalizes the one we saw in equation \eqref{Hi} for a pointlike detector. Indeed, notice that if $f(\bm\xi)=\delta^{(d)}(\bm \xi)$ (if we consider a pointlike detector), \eqref{sdet0} becomes exactly \eqref{Hi} in the interaction picture.

For this smeared detector, the switching and the smearing are prescribed in the centre of mass reference frame of the detector. General coordinate transformations will, however, mix time and space. To properly perform coordinate transformations it is convenient to construct an Unruh-DeWitt  Hamiltonian density: 
\begin{equation}\label{sdetM1}
    _{D}\hat{h}_{I}^{\tau}\coloneqq \hbar c\lambda \chi(\tau) f(\bm \xi)\hat\mu(\tau)\hat \phi[t(\tau,\bm \xi),\bm x(\tau,\bm\xi)],
\end{equation}
so that the Hamiltonian in this frame is
\begin{equation}\label{sdetM2}
    _{D}\hat{H}_{I}^{\tau}=\int \dd^d \bm\xi \, _{D}\hat{h}_{I}^{\tau}.
\end{equation}

This is convenient because the time evolution operator is the time-ordered exponential of the Hamiltonian integrated in time, which is a space-time integral of the Hamiltonian density
\begin{equation}
    \hat U=\mathcal{T}\exp\left(\frac{-\ii}{\hbar}\int \dd^n \bm \xi\,\dd \tau\,_{D}\hat{h}_{I}^{\tau}\right).
\end{equation}
Note, that the integral is taken over the whole spacetime, and this evolution operator takes the state of the detector and field on a  Cauchy surface in the asymptotic past, and evolves it to a Cauchy surface in the asymptotic future. The localized nature of the interaction and its finite duration in time comes through the time and spatial support of the switching and smearing functions in the Hamiltonian density.  We can find the Hamiltonian density in a different reference frame precisely demanding that time evolution is invariant under changes of frame. For example, consider a different frame $(t,\bm x)$. The Hamiltonian density generating time translations in that frame $_{D}\hat{h}_{I}^{t}$ is fixed demanding that 
\begin{align}
    \hat U=\mathcal{T}&\exp\!\left(\frac{-\ii}{\hbar}\int\! \dd^n \bm \xi\,\dd \tau\,_{D}\hat{h}_{I}^{\tau}\right)\\
    &=\mathcal{T}\exp\left(\frac{-\ii}{\hbar}\int \dd^n \bm x\,\dd t\,_{D}\hat{h}_{I}^{t}\right).\nonumber
\end{align}
In particular, if the frame $(t,\bm x)$ corresponds to the quantization inertial frame of the field, the Hamiltonian density in the  frame  $(t,\bm x)$ is
\begin{equation}\label{sdetM3}
    _{D}\hat{h}_{I}^{t}= \hbar c\lambda \chi[\tau(t,\bm x)] f[\bm \xi(t,\bm x)]\hat\mu[\tau(t,\bm x)]\hat \phi(t,\bm x)\left|\frac{\partial (\tau,\bm\xi)}{\partial (t,\bm x)}\right|,
\end{equation}
where $\left|\frac{\partial (\tau,\bm\xi)}{\partial (t,\bm x)}\right|$ is the Jacobian of the change of coordinates.

Therefore, the transformed Hamiltonian that generates translations with respect to the lab time $t$ is the integral over space of the corresponding Hamiltonian density. Namely,
\begin{align}\label{sdeta}
&_{D}\hat{H}_{I}^{t}=\int \dd^d\bm x\, _{D}\hat{h}_{I}^{t} \\&=\hbar c\lambda \int\dd^d\bm x\, \chi[\tau(t,\bm x)] f[\bm \xi(t,\bm x)]\nonumber\hat\mu[\tau(t,\bm x)]\hat \phi(t,\bm x)\left|\frac{\partial (\tau,\bm\xi)}{\partial (t,\bm x)}\right|.
\end{align}
Note that, in general, it is not possible to write the Hamiltonian as a switching function times an integral over a (time independent) smeared field observable in a different frame than originally prescribed. Instead time and space get obviously mixed in the new Hamiltonian.

A particularly illustrative example that will be useful for later discussions is to see what is the interaction Hamiltonian from the lab frame for a  detector that moves in an inertial trajectory with respect to the lab frame. Without loss of generality, for this example we can choose the velocity to be in the $x$ direction so that the trajectory is given by $(x(t),y(t),z(t))=(v t ,0 ,0)$. The coordinate transformation between the centre of mass frame and the lab frame is a simple Lorentz transformation
 \begin{equation}
    \tau(t, x)=\gamma \left(t- \frac{x v}{c^2}\right),\qquad  \xi_1(t, \bm x)=\gamma( x -  vt),
\end{equation}
together with $\xi_2=y, \xi_3=z$,  where $\gamma=(1-\frac{v^2}{c^2})^{-1/2}$. The Jacobian of a Lorentz transformation is one. Indeed: 
\begin{equation}
\begin{vmatrix}
\partial_t\tau & \partial_{x}\tau \\
\partial_t\xi & \partial_{x}\xi
\end{vmatrix}=\begin{vmatrix}
\gamma  & \gamma\frac{v}{c^2} \\
\gamma v & \gamma
\end{vmatrix}=\gamma^2\left(1-\frac{v^2}{c^2}\right)=1;
\end{equation}
substituting this in \eqref{sdeta} we get
\begin{align}\label{sdetampa}
    _{D}\hat{H}_{I}^{t}&=\hbar c\lambda \int\dd^d\bm x\, \chi\left[\gamma \left(t- \frac{ x v}{c^2}\right)\right] f\left[\gamma(\bm x -  v t),y,z\right]\nonumber\\
    & \times\hat\mu\left[\gamma \left(t- \frac{\bm x v}{c^2}\right)\right]\hat \phi(t,\bm x).
\end{align}
If $f(\bm \xi)=\delta^{(d)}(\bm\xi)$ then the spatial integral can be performed, and its effect is that it  imposes the constraint that precisely the field is evaluated along the trajectory of the detector $\bm x=\bm v t$, using that (as a distribution) $\delta (a x)=\frac1a\delta(x)$ the spatial integration yields
\begin{equation}\label{sdelta}
    _{D}\hat{H}_{I}^{t}=\hbar c\lambda\gamma^{-1}  \chi\left[\gamma^{-1} t\right] \hat\mu\left(\gamma^{-1} t\right)\hat \phi(t,\bm v t)
\end{equation}
for the Hamiltonian that generates translations from the lab frame in the case of a pointlike detector on an inertial trajectory.

One can wonder what would be the Hamiltonian if the switching function is controlled from the lab frame, that is, the experimenter is the one switching on and off the interaction with the field. For example one can think of a device that prepares the state of an atom at the entrance of an optical cavity. The device would carry out a projective measurement on the atom in its free energy eigenbasis such that it ends up prepared in the ground state (i.e. we measure with optical means, and post select only on ground states). Then the atom enters the cavity transversely (or forming some angle with the axis of the mirrors) spending a finite time interacting with the field. Or for example, in microwave cavities with superconducting qubits, it is possible to control the strength of the coupling of a superconducting qubit with the quantum electromagnetic field inside a microwave guide as a function of time, and to make it vary either smoothly or sharply following any desired profile (see e.g.~\cite{Peropadre2010,Fornada}).

The temptation in this case could be to just write that, in the lab frame, we can factor  the switching function out of the integral in space. We will check whether that expectation is well aimed or not.

The Hamiltonian density in this case would be given by
\begin{equation}\label{EduMea}
_{D}\hat{h}_{I}^{\tau} = \hbar c\lambda \chi(t(\tau,\bm \xi)) f(\bm \xi)\hat\mu(\tau)\hat \phi[t(\tau,\bm \xi),\bm x(\tau,\bm\xi)].
\end{equation}
Notice that, unsurprisingly, if it is the laboratory frame that switches the interaction, different points of the detector will perceive the switching of the interaction in nonsimultaneous instants. In the same fashion as above, we can now readily transform \eqref{EduMea} to obtain the Hamiltonian density that will generate transformations with respect to lab time:
\begin{equation}
_{D}\hat{h}_{I}^{t}  =  \hbar c\lambda \chi(t) f(\bm{\xi}(t,\bm{x}))\hat\mu(\tau(t,\bm{x}))\hat{\phi}[t,\bm{x}]
\left|\frac{\partial (\tau,\bm\xi)}{\partial (t,\bm x)}\right|.
\end{equation}
From here we can obtain the Hamiltonian integrating the Hamiltonian density in space:
\begin{equation}
_{D}\hat{H}_{I}^{t} = \hbar c\lambda \chi(t)\int\dd^{d}\bm{x} f(\bm{\xi}(t,\bm{x}))\hat\mu(\tau(t,\bm{x}))\hat{\phi}[t,\bm{x}]
\left|\frac{\partial (\tau,\bm\xi)}{\partial (t,\bm x)}\right|,
\end{equation}
where we see that even though the switching function $\chi(t)$ factors out there is also extra time dependence in  the smearing function.

\section{The simple case of pointlike detectors in arbitrary trajectories}

For a pointlike detector $f(\bm \xi)=\delta^{(d)}(\bm \xi)$, carrying out the spatial integral in \eqref{sdet0} reduces the Hamiltonian to the monopole moment of the detector coupled to the pull-back of the field on the detector's trajectory. Indeed, upon integration over $\bm \xi$, \eqref{sdet0}  becomes
\begin{equation}\label{famous}
    _{D}\hat{H}_{I}^{\tau}= \hbar c\lambda \chi(\tau)\hat\mu(\tau)\hat \phi[t(\tau),\bm x(\tau)],
\end{equation}
where $t(\tau)=t(\tau,\bm 0)$, $\bm x(\tau)=\bm x(\tau,\bm 0)$ is the trajectory of the centre of mass of the atom from the frame $(t,\bm x)$. This is the standard form of the Unruh-DeWitt Hamiltonian as originally introduced by DeWitt and as is most commonly used in the literature.

In this simple case, the transformation of the Hamiltonian generating translations with respect to $\tau$ to the Hamiltonian generating translations with respect to $t$ is reduced to a time reparametrization. Indeed, for the interaction picture time evolution to be invariant under change of frame, we need to demand
\begin{align}\label{reparampointlike}
    \hat U &=\mathcal{T}\exp\left[\frac{-\ii}{\hbar}{\int_{\mathbb{R}} \dd \tau\,_{D}\hat{H}_{I}^{\tau}}\right]=\mathcal{T}\exp{\left[\frac{-\ii}{\hbar}\int_{\mathbb{R}} \dd t\,_{D}\hat{H}_{I}^{t}(t)\right]},
\end{align}
which, completely analogously to \eqref{reparam}, tells us that for the pointlike case 
\begin{equation}\label{anotheronebitesthedust}
_{D}\hat{H}_{I}^{t}(t)=\frac{\dd t}{\dd \tau} {_{D}\hat{H}_{I}^{\tau}[\tau(t)]}
\end{equation}

Let us consider that the detector moves along  some parametric timelike curve $\bm{x}(t)$ as seen in the lab frame (an inertial frame we choose to do the quantization of the field). The relationship between the proper time of the detector $\tau$ and the lab coordinate time $t$ can be trivially worked out:
\begin{eqnarray}\label{gamma1}
\dd s^{2} & = & - c^{2}\dd\tau = - c^{2}\dd t^{2} + \dd\bm{x}^{2}\nonumber\\
& \Rightarrow & \frac{\dd\tau}{\dd t} =\sqrt{1-\frac{1}{c^{2}}\left(\frac{\dd\bm{x}}{\dd t}\right)^{2}}\nonumber\\
& \Rightarrow & \dd\tau=[\gamma(t)]^{-1}\dd t,
\end{eqnarray}
where
\begin{equation}\label{gamma2}
    \gamma(t)\coloneqq\frac{1}{\sqrt{1-\left(\dfrac{\bm v(t)}{c}\right)^{2}}}
\end{equation}
and $\bm v(t)\coloneqq \dfrac{\dd\bm{x}}{\dd t}$ is the velocity of the particle at the instant $t$ measured in the lab frame.

Therefore we can write the relationship between the two times as a total integral
\begin{equation}\label{A7}
    \tau(t)=\int^t\frac{\dd t'}{\gamma(t')},  
\end{equation}
together with a matching condition [for example that $\bm v(t_0)=0$, or any other]. Eq. \eqref{A7} implicitly defines the function $t(\tau)$, obtained taking its inverse.  

Note that if one knows the function $t(\tau)$ by inversion of \eqref{A7}, or alternatively if one has the knowledge of the trajectory in the inertial frame parametrized in terms of proper time $(t(\tau),\bm{x} (\tau))$ one can compute $ \bm v[t(\tau)]$ and write the inverse relation as
\begin{equation}\label{tdetau}
    t(\tau)=\int^\tau \dd \tau' \gamma(\tau').
\end{equation}
Therefore, from equation \eqref{anotheronebitesthedust}, we see that the Hamiltonian $_{D}\hat{H}_{I}^{t}(t)$ that generates translations with respect to the lab frame $t$ can be rewritten in terms of the Hamiltonian  $_{D}\hat{H}_{I}^{\tau}(\tau)$ that generates translations with respect to the detector's proper time $\tau$ as 
\begin{align}\label{startingpoint}
\nonumber _{D}\hat{H}_{I}^{t}(t)=&[\gamma(t)]^{-1} {_{D}\hat{H}_{I}^{\tau}[\tau(t)]}\\&= \hbar c \lambda [\gamma(t)]^{-1} \chi[\tau(t)]\hat\mu[\tau (t)]\hat \phi(t,\bm x).
\end{align}
Note how, as anticipated earlier, in the case that the detector moves initially, \eqref{startingpoint} simply becomes \eqref{sdelta}.

\section{Examples: Invariance of the transition probability}

We have shown how to obtain the form of the interaction Hamiltonian in different reference frames, and in particular, in the cases of the detector's proper frame and the laboratory frame (an inertial frame where we perform the field quantization). All the results above are nonperturbative in nature, however, as an illustrative example, we can compute explicitly the leading order contribution to the vacuum excitation probability for an atom initially in the ground state, as well as the probability for spontaneous emission if the atom starts in the excited state.

\subsection{Pointlike detector vacuum excitation calculation in the detector's frame}

We start from the Hamiltonian \eqref{famous}, after expanding the field in an orthonormal basis of plane wave modes
\begin{eqnarray}
_{D}\hat{H}_{I}^{\tau} & = & \hbar c\lambda\chi(\tau)
\left( \hat{\sigma}^{+}e^{\ii\Omega\tau} + \hat{\sigma}^{-} e^{-\ii \Omega \tau} \right)
\int\frac{\dd^{d}\bm{k}}{\sqrt{2(2\pi)^{d} \omega_{\bm{k}}c^{-1}}}\nonumber\\
& & \!\!\!\times\left[ 
   \hat{a}_{ \bm{k}}^{\dagger}e^{\ii[  \omega_{\bm{k}}t(\tau)-\bm{k} \cdot \bm{x}(\tau)]}
 + \hat{a}_{ \bm{k}}e^{-\ii [\omega_{\bm{k}}t(\tau)-\bm{k} \cdot \bm{x}(\tau)]}
\right].
\end{eqnarray}
Since we are interested in models for the light-matter interaction, in the following we will particularize to the massless case, where $\omega_{\bm{k}} = c\abs{\bm{k}}$. We can proceed to compute the transition probability assuming that the initial state of the field and detector is the ground state $\ket{g,0}$. Using Born's rule 
\begin{equation}\label{jorona}
    P(\Omega)=\sum_{\text{out}}\left|\bra{e,\text{out}}\hat U \ket{g,0}\right|^{2},
\end{equation}
where the sum over states $\ket{\text{out}}$ represent a sum over an orthonormal basis of possible final states of the field. The time evolution operator in the interaction picture is
\begin{equation}
    \hat U =\mathcal{T}\,\,\,\text{exp}\left( - \frac{\ii}{\hbar}\int_{-\infty}^{\infty}\dd\tau \hat{H}_{\II}(\tau) \right),
\end{equation}
being $\mathcal{T}$ the time ordering operator. Taking a Dyson expansion
\begin{equation}
\hat U = \openone + \hat U^{(1)} + \mathcal{O}\left(\lambda^{2}\right),
\end{equation}
where $
\hat U^{(1)} = - \frac{\ii}{\hbar} \int_{-\infty}^{\infty}\dd\tau \hat H_{\II}(\tau)$.
The leading order contribution to the transition probability is given by
\begin{eqnarray}
P(\Omega) 
 & = & \sum_{\text{out}} \bra{g,0} (\hat U^{(1)})^{\dagger} \ket{e,\text{out}} \bra{e,\text{out}}\hat U^{(1)} \ket{g,0}\nonumber\\
& & + \mathcal{O}(\lambda^{4}).
\end{eqnarray}
Note that the next subleading order is $\lambda^{4}$ because the third order correction to the probability cancels. Substituting the interaction Hamiltonian \eqref{famous} we get that
\begin{align}\label{intermed0}
P(\Omega) &= c^{2}\lambda^{2}\sum_{\text{out}} \int_{-\infty}^\infty \!\!\!\!\!\dd\tau \int_{-\infty}^\infty \!\!\!\!\!\dd\tau' \chi(\tau)\chi(\tau')\\
& \times\Big[\bra{g}\hat\mu(\tau)  \ket{e} \bra{e}\hat\mu(\tau')\ket{g}\nonumber\\
& \times\bra{0} \hat\phi[t(\tau),\bm{x}(\tau)] \ket{\text{out}}\!\bra{\text{out}}\hat\phi[t(\tau'),\bm{x}(\tau')] \ket{0} \Big]\nonumber\\
& + \mathcal{O}(\lambda^{4}).\nonumber
\end{align} 
Using that $\sum_{out}\proj{\text{out}}{\text{out}}=\openone$ and that 
\begin{equation}
    \bra{g}\hat\mu(\tau)  \ket{e} \bra{e}\hat\mu(\tau')  \ket{g}=e^{-\ii\Omega (\tau -\tau')},
\end{equation}
we can write \eqref{intermed0} as
\begin{eqnarray}\label{intermed2}
P(\Omega) & = & c^{2}\lambda^{2}\int_{-\infty}^{\infty} \!\!\!\!\!\dd\tau \int_{-\infty}^\infty \!\!\!\!\!\dd\tau' \chi(\tau)\chi(\tau') e^{-\ii\Omega (\tau -\tau')}\nonumber\\
    & & \times W[t(\tau),\bm{x}(\tau),t(\tau'),\bm{x}(\tau')] + \mathcal{O}(\lambda^{4}),
\end{eqnarray}
where the vacuum Wightman function is
\begin{align}
& W[t(\tau),\bm{x}(\tau),t(\tau'),\bm{x}(\tau')] \\
& =\bra{0} \hat\phi[t(\tau),\bm{x}(\tau)]\hat\phi[t(\tau'),\bm{x}(\tau')]\ket{0}\nonumber\\
& =\int  \frac{\dd^{d}\bm{k}}{2(2\pi)^{d}\abs{\bm{k}}} 
e^{-\ii[  c\abs{\bm{k}}(t(\tau)-t(\tau')) -\bm{k} \cdot (\bm{x}(\tau)-\bm{x}(\tau')]}.\nonumber
\end{align}
Upon substitution of the Wightman function in \eqref{intermed2} the transition probability can be simplified to
\begin{eqnarray}\label{intermed3}
P(\Omega) & = & c^{2}\lambda^{2}\int\frac{\dd^{d}\bm{k}}{2(2\pi)^{d}\abs{\bm{k}}}\\
& & \times\left| \int_{-\infty}^\infty \!\!\!\!\!\dd\tau\,  \chi(\tau) e^{-\ii[\Omega \tau + c\abs{\bm{k}}t(\tau)-\bm{k}\cdot\bm{x}(\tau)]} \right|^{2} + \mathcal{O}(\lambda^{4}).\nonumber
\end{eqnarray}
\subsubsection*{Particularizing to the inertial case}
In the particular case where the trajectory of the detector is inertial with some nonzero speed with respect to the lab frame, we get that
\begin{equation}\label{traject0}
\bm{x}(t)=\bm v t \Rightarrow \left\{\begin{array}{ll}
\tau(t)=\gamma(t-\frac{\bm v\cdot\bm{x}}{c^2})\Rightarrow t(\tau)=\gamma\tau\\
\bm{x}(\tau)=\bm v t(\tau)=\gamma \bm v t,
\end{array}\right.
\end{equation}
where in this case $\gamma=\left(1-\frac{\bm v^{2}}{c^{2}}\right)^{-\frac12}$ is constant. In this case, the transition probability takes the simple form
\begin{align}\label{intermed3final}
P(\Omega) & = c^{2}\lambda^{2}\int\frac{\dd^{d}\bm{k}}{2(2\pi)^{d}\abs{\bm{k}}}\\
& \times\left| \int_{-\infty}^\infty \!\!\!\!\!\dd\tau\,  \chi(\tau) e^{-\ii(\Omega  + c\gamma\abs{\bm{k}}-\gamma\bm{k}\cdot\bm v)\tau} \right|^{2} + \mathcal{O}(\lambda^{4})\nonumber\\
& \hspace{-1cm}= c^{2}\lambda^{2}\int\frac{\dd^{d}\bm{k}}{2(2\pi)^{d} \abs{\bm{k}}}\nonumber
\left|   \bar\chi(\Omega  + c\gamma\abs{\bm{k}}-\gamma\bm{k}\cdot\bm v) \right|^{2} + \mathcal{O}(\lambda^{4}),
\end{align}
where  $
 \bar\chi(\Omega)\coloneqq\int \!\dd \tau \,\chi(\tau)e^{\ii \Omega \tau},$ 
is the Fourier transform of the switching function. Notice that the transition probability starting from the exciting state instead of the ground state is obtained swapping $\Omega\rightarrow -\Omega$.

\subsection{Pointlike detector vacuum excitation calculation in the lab frame}

Starting from \eqref{startingpoint}, and particularizing for a massless field, we can write the interaction Hamiltonian in the interaction picture generating translations with respect to $t$ as
\begin{align}\label{calculacabrula}
_{D}\hat{H}_{I}^{t} & =  
\hbar c\lambda\frac{\chi[\tau(t)]}{{\gamma(t)}}\left( \hat{\sigma}^{+}e^{\ii\Omega\tau(t)} + \hat{\sigma}^{-}e^{-\ii\Omega\tau(t)} \right)\nonumber\\
& \times
\int\frac{\dd^{d}\bm{k}}{\sqrt{2(2\pi)^{d} \abs{\bm{k}}}}\\
& \quad \times\left[ 
   \hat{a}_{ \bm{k}}^{\dagger}e^{\ii[  c\abs{\bm{k}}t-\bm{k} \cdot \bm{x}(t)]}
 + \hat{a}_{ \bm{k}}e^{-\ii [c\abs{\bm{k}}t-\bm{k} \cdot \bm{x}(t)]}
\right]\nonumber
\end{align}

Now we can proceed again to compute the transition probability assuming that the initial state of the field and detector is the ground state $\ket{g,0}$. Using Born's rule and repeating the steps in the previous section but replacing the interaction Hamiltonian with \eqref{calculacabrula}, we arrive to
\begin{align}\label{intermed}
P(\Omega) &=c^{2}\lambda^{2}\sum_{\text{out}} \int_{-\infty}^\infty \!\!\!\!\!\dd t \int_{-\infty}^\infty \!\!\!\!\!\dd t' \frac{\chi[\tau(t)]\chi[\tau(t')]}{\gamma(t)\gamma(t')}  \\
& \times\Big[\bra{g}\hat\mu[\tau(t)]  \ket{e} \bra{e}\hat\mu[\tau(t')]  \ket{g}\nonumber\\
& \times\bra{0} \hat\phi[t,\bm{x}(t)] \ket{\text{out}}\!\bra{\text{out}}\hat\phi[t',\bm{x}(t')] \ket{0} \Big] + \mathcal{O}(\lambda^{4}).\nonumber
\end{align}
Same as above, using that $\sum_{out}\proj{\text{out}}{\text{out}}=\openone$ and that $
    \bra{g}\hat\mu(\tau)  \ket{e} \bra{e}\hat\mu(\tau')  \ket{g}=e^{-\ii\Omega (\tau -\tau')}$
we can write \eqref{intermed} as
\begin{eqnarray}\label{intermed24}
P(\Omega) & = & c^{2}\lambda^{2}\int_{-\infty}^\infty \!\!\!\!\!\dd t \int_{-\infty}^\infty \!\!\!\!\!\dd t' \frac{\chi[\tau(t)]\chi[\tau(t')]}{\gamma(t)\gamma(t')} e^{-\ii\Omega [\tau(t) -\tau(t')]}\nonumber\\
& & \times W[t,\bm{x}(t),t',\bm{x}(t')] + \mathcal{O}(\lambda^{4}).
\end{eqnarray} 
where the vacuum Wightman function is
\begin{align}
W[t,\bm{x}(t),t',\bm{x}(t')] & = \bra{0} \hat\phi[t,\bm{x}(t)]\hat\phi[t',\bm{x}(t')]\ket{0}\\
& \hspace{-1cm}= \int\frac{\dd^{d}\bm{k}}{2(2\pi)^{d} \abs{\bm{k}}} 
e^{-\ii[  c\abs{\bm{k}}(t-t') -\bm{k} \cdot (\bm{x}(t)-\bm{x}(t')]}.\nonumber
\end{align}
Upon substitution of the Wightman function in \eqref{intermed24} the transition probability can be simplified as
\begin{eqnarray}\label{intermed34}
P(\Omega) & = & c^{2}\lambda^{2}\int\frac{\dd^{d}\bm{k}}{2(2\pi)^{d}\abs{\bm{k}}}\\
& \times & \left| \int_{-\infty}^\infty\! \frac{\dd t}{\gamma(t)}  \chi[\tau(t)] e^{-\ii[\Omega \tau(t) + c\abs{\bm{k}}t-\bm{k}\cdot\bm{x}(t)]} \right|^{2} + \mathcal{O}(\lambda^{4}).\nonumber
\end{eqnarray} 
On this expression we can perform the change of variables 
\begin{equation}\label{changeov}
    \frac{\dd t}{\gamma(t)}=\dd\eta \Rightarrow \eta= \int^t \frac{\dd t}{\gamma(t)} \Rightarrow t\rightarrow t(\eta).
\end{equation}
Notice that the dummy variable $\eta$ plays the same role as $\tau$ and $t(\eta)$ plays the same role as $t(\tau)$ as given by expressions \eqref{gamma1}, \eqref{A7}. Therefore we see, as announced, that in this general trajectory pointlike case, \eqref{intermed34} yields exactly the same exact expression as in the detector frame calculation \eqref{intermed3}.

\subsubsection*{Particularizing to the inertial case}
To illustrate the invariance of the leading order transition probability in an even clearer particular example, let us evaluate  explicitly for the inertial case  the probability \eqref{intermed34}: In the particular case where the trajectory of the detector is inertial following the trajectory \eqref{traject0} we get that
\begin{equation}\label{traject02}
    \bm{x}(t)=\bm v t,\qquad  
    \tau(t)=\gamma\left(t-\frac{\bm v\cdot\bm{x}}{c^2}\right)= \gamma^{-1}t,
\end{equation}
where in this case $\gamma=\left(1-\frac{\bm v^{2}}{c^{2}}\right)^{-\frac12}$ is constant. In this case, the transition probability \eqref{intermed34} takes the simple form
\begin{eqnarray}\label{intermed34final1}
P(\Omega) & & =c^{2}\lambda^{2}\gamma^{-2}\int\frac{\dd^{d}\bm{k}}{2(2\pi)^{d} \abs{\bm{k}}}\\
& \times & \left| \int_{-\infty}^\infty\!\dd t\,  \chi[\gamma^{-1} t] e^{-\ii[\gamma^{-1}\Omega t + c\abs{\bm{k}}t-\bm{k}\cdot\bm v t]} \right|^{2} + \mathcal{O}(\lambda^{4}).\nonumber
\end{eqnarray} 
Multiplying and dividing the exponent in the integrand by $\gamma$ we get
\begin{eqnarray}\label{intermed34final2}
P(\Omega) & = & c^{2}\lambda^{2}\gamma^{-2}\int\frac{\dd^{d}\bm{k}}{2(2\pi)^{d}\abs{\bm{k}}}\\
& \times & \left| \int_{-\infty}^\infty\!\dd t \, \chi[\gamma^{-1} t] e^{-\ii\gamma^{-1} t[\Omega  + c\gamma\abs{\bm{k}}-\gamma\bm{k}\cdot\bm v ]} \right|^{2} + \mathcal{O}(\lambda^{4})\nonumber
\end{eqnarray}
Finally, performing the change of variables suggested in \eqref{changeov}, which in this simple case is
\begin{equation}
    \eta=\gamma^{-1}t\Rightarrow \dd \eta =\gamma^{-1}\dd t
\end{equation}
\eqref{intermed34final2} yields
\begin{eqnarray}\label{intermed3final3}
P(\Omega) & = & c^{2}\lambda^{2}\int\frac{\dd^{d}\bm{k}}{2(2\pi)^{d} \abs{\bm{k}}}\\
& \times & \left| \int_{-\infty}^\infty \!\!\!\!\!\dd\eta\,  \chi(\eta) e^{-\ii(\Omega  + c\gamma\abs{\bm{k}}-\gamma\bm{k}\cdot\bm v)\eta} \right|^{2} + \mathcal{O}(\lambda^{4})\nonumber\\
& = & c^{2}\lambda^{2}\int\frac{\dd^{d}\bm{k}}{2(2\pi)^{d} \abs{\bm{k}}}\nonumber\\
& \times & \left| \int_{-\infty}^\infty \!\!\!\!\!\dd\tau\,  \bar\chi(\Omega  + c\gamma\abs{\bm{k}}-\gamma\bm{k}\cdot\bm v) \right|^{2} + \mathcal{O}(\lambda^{4}),\nonumber
\end{eqnarray}
where $\bar\chi(\Omega)$ is the Fourier transform of the switching function. This is exactly the same transition probability obtained in the detector frame calculation \eqref{intermed3final}.

\subsection{Vacuum transition probability for an inertial smeared detector}

We start now from the smeared detector Hamiltonian \eqref{sdet0} and we proceed to compute the transition probability assuming that the initial state of the field and detector is the ground state $\ket{g,0}$. Same as in the previous sections, using Born's rule we get
\begin{equation}
    P(\Omega)=\sum_{\text{out}}\left|\bra{e,\text{out}}\hat U \ket{g,0}\right|^{2}=P^{(2)}+\mathcal{O}(\lambda^4)
\end{equation}

Repeating a calculation completely analogous to the one leading from \eqref{jorona} to \eqref{intermed2}, we obtain for the smeared case
\begin{align}\label{intermed2s}
P^{(2)} & =  c^{2}\lambda^{2}\int_{-\infty}^{\infty} \!\!\!\!\!\dd\tau \int_{-\infty}^\infty \!\!\!\!\!\dd\tau'\int \!\dd^n\bm\xi\int \!\dd^n\bm\xi' \chi(\tau)\chi(\tau')f(\bm\xi)  \nonumber\\
    &\!\!\!\!\!\!\!\! \times f(\bm\xi')e^{-\ii\Omega (\tau -\tau')} W[t(\tau,\bm \xi),\bm{x}(\tau,\bm \xi),t(\tau',\bm \xi'),\bm{x}(\tau',\bm \xi')] 
\end{align}
where the Wightman function is
\begin{align}\label{Wiggy}
& W[t(\tau,\bm \xi),\bm{x}(\tau,\bm \xi),t(\tau',\bm \xi'),\bm{x}(\tau',\bm \xi')]  \\
& =\bra{0} \hat\phi[t(\tau,\bm \xi),\bm x(\tau,\bm\xi)]\hat\phi[t(\tau',\bm \xi'),\bm x(\tau',\bm\xi')]\ket{0}\nonumber\\
& =\int  \frac{\dd^{d}\bm{k}}{2(2\pi)^{d}\abs{\bm{k}}} 
e^{-\ii[  c\abs{\bm{k}}(t(\xi,\tau)-t(\xi',\tau')) -\bm{k} \cdot (\bm{x}(\bm \xi,\tau)-\bm{x}(\bm \xi',\tau'))]}.\nonumber
\end{align}

Let us consider for illustration the problem of computing the transition probability of a smeared atom whose centre of mass undergoes an inertial trajectory parametrized in the lab frame as $\bm x(t)=\bm v t$.

In that case, $t(\tau, \bm\xi)$ and $\bm x(\tau, \bm\xi)$ are given by simple Lorentz transformations:
\begin{equation}\label{Lcor}
    t(\tau, \bm\xi)=\gamma \left(\tau-\xi_\parallel \frac{|\bm v|}{c^2}\right),\quad \bm x(\tau, \bm\xi)=\gamma(\xi_\parallel -|\bm v|\tau)\frac{\bm v}{|\bm v|}+\bm \xi_\perp
\end{equation}
where we have decomposed $\bm \xi$ in components parallel and perpendicular to the velocity of the atom with respect to the lab frame, i.e., $\xi_\parallel=\bm\xi\cdot\frac{\bm v}{|\bm v|}$ and $\bm \xi =\xi_\parallel\frac{\bm v}{|\bm v|}+\bm \xi_\perp.$

Substituting the change of coordinates \eqref{Lcor} [corresponding to the atomic trajectory $\bm x(t)$] into \eqref{Wiggy} we obtain
\begin{align}\label{Wiggy2}
\nonumber W[t(\tau,\bm \xi),&\bm{x}(\tau,\bm \xi),t(\tau',\bm \xi'),\bm{x}(\tau',\bm \xi')]  \\\nonumber
 =\int & \frac{\dd^{d}\bm{k}}{2(2\pi)^{d}\abs{\bm{k}}} 
e^{\ii[  \abs{\bm{k}}\gamma[c(\tau'-\tau)-\frac{\bm v}{c}\cdot(\bm\xi'-\bm\xi) ]} \\
\nonumber&\times e^{-\ii k_{\parallel}\gamma[\xi'_\parallel-\xi_\parallel-|\bm v|(\tau'-\tau)]}e^{-\ii\bm k_\perp\cdot(\bm\xi'_\perp-\bm\xi_\perp)}\\
&=\int  \frac{\dd^{d}\bm{k}}{2(2\pi)^{d}\abs{\bm{k}}} 
e^{-\ii[c\tilde k_0 (\tau'-\tau)+\tilde{\bm k}\cdot(\bm \xi'-\bm\xi)]}.
\end{align}
where $k_\parallel=\bm k\cdot\frac{\bm v}{|\bm v|}$ and $\bm k_\perp=\bm k- k_\parallel\frac{\bm v}{|\bm v|}$, and in the last step  we have defined
\begin{align}
   \label{k0tilda} \tilde{k}_{0} &\coloneqq \gamma\left(-|\bm k|+\frac{\bm k\!\cdot\!\bm v}{c}\right)\\
    \tilde{\bm k} &\coloneqq \gamma\left(|\bm k|\frac{\bm v}{c}+\bm k\!\cdot\!\bm v\frac{\bm v}{|\bm v|^2}\right)+\bm k_\perp 
\end{align}
so that we could write the exponent of \eqref{Wiggy2} in the convenient form
\begin{equation}
    \tilde k_\mu (\xi'^\mu-\xi^\mu)\coloneqq c\tilde k_0 (\tau'-\tau)+\tilde{\bm k}\cdot(\bm \xi'-\bm\xi ). 
\end{equation}

Substituting the final expression of the Wightman function \eqref{Wiggy2} into the transition probability \eqref{intermed2s} we see that the integrals over $\bm \xi$ and $\bm \xi'$ become Fourier transforms of the smearing functions, explicitly:
\begin{align}\label{intermed3s}
P^{(2)} & = c^{2}\lambda^{2}\int\frac{\dd^{d}\bm{k}}{2\abs{\bm{k}}}\frac{|\bar f(\tilde{\bm k})|^2}{(2\pi)^{d}}\nonumber\\
& \times\int_{-\infty}^{\infty}\!\!\!\dd\tau\int_{-\infty}^{\infty}\!\!\!\dd\tau'
\chi(\tau)\chi(\tau') e^{\ii(\Omega-\tilde k_0)(\tau'-\tau)} 
\end{align}
where
\begin{equation}
 \bar f(\bm q)\coloneqq\int \!\dd^n\bm\xi \,f(\bm \xi)e^{\ii\bm q\cdot \bm xi}.
\end{equation}
Finally,\eqref{intermed3s} can be further simplified in terms of the Fourier transform of the switching function:
\begin{align}\label{finalS1}
P^{(2)} & = c^{2}\lambda^{2}\!\!\!\int\!\frac{\dd^{d}\bm{k}}{2(2\pi)^{d}\abs{\bm{k}}}|\bar f(\tilde{\bm k})|^2 |\bar\chi(\Omega-\tilde k_0)|^2. 
\end{align}
Notice that particularizing for a pointlike detector $f(\bm \xi)=\delta^{(d)}(\bm \xi)$ reproduces the result \eqref{intermed3final} in the pointlike section. Also, note that for $\bm v=\bm 0\Rightarrow \gamma=1$ and from \eqref{k0tilda} $\Omega-\tilde k_0$ becomes $\Omega+|\bm k|$.

\section{conclusions}

In this note we have discussed how the Hamiltonian of particle detectors  coupling degrees of freedom of first quantized systems to quantum fields (and in particular the Unruh-DeWitt detector) transform under general changes of coordinates. 

The aim of this note was to pedagogically show the form of such Hamiltonians in different frames for particle detectors that may move in relativistic general trajectories with respect to the lab frame, as well as to show that there is no problem associated with the relativistic invariance of the predictions of these models. This is particularly relevant in studies in relativistic quantum information where detectors move in arbitrary trajectories with arbitrary switching functions.  

More specifically, we have discussed what is a reasonable prescription for the interaction Hamiltonian of smeared detectors (such as e.g.,  hydrogenoid atoms) and we have shown the general form of the interaction Hamiltonian under changes of reference frame, paying special attention to the effect on the switching and the smearing functions of the mix of time and space in the coordinate transformation. For illustration, and for didactic purposes, we have also particularized to the common case of pointlike detectors showing a specific calculation of the vacuum excitation probability as well as the spontaneous emission probability to show explicitly the invariance under changes of reference frame.

Finally, note that while we have worked in flat space in this paper, obtaining the same results in curved spaces is not any more involved, and should be straightforward from the work in the paper introducing the determinant of the metric and the metric coefficients where it corresponds, but the procedural details are the same.

\section{Acknowledgment}

The authors would like to very much thank Luis J. Garay for immensely helpful discussions on the importance of the invariance of the evolution operator under general transformations. These authors are indebted to Luis in much more than what a article can capture. E.M.-M. is thankful for helpful discussions with Allison Sachs, Laura Henderson and Robert B. Mann. E.M.-M. acknowledges the support of the NSERC Discovery program as well as the Ontario Early Researcher Award.
P.R.-L. also acknowledges partial support from TerMic (Grant No. FIS2014-52486-R, Spanish Government), CONTRACT (Grant No. FIS2017-83709-R, Spanish Government) and from Juan de la Cierva - Incorporacion program (Ref: I JCI-2015-25315, Spanish Government).

\bibliography{references}

\begin{thebibliography}{54}%
\makeatletter
\providecommand \@ifxundefined [1]{%
 \@ifx{#1\undefined}
}%
\providecommand \@ifnum [1]{%
 \ifnum #1\expandafter \@firstoftwo
 \else \expandafter \@secondoftwo
 \fi
}%
\providecommand \@ifx [1]{%
 \ifx #1\expandafter \@firstoftwo
 \else \expandafter \@secondoftwo
 \fi
}%
\providecommand \natexlab [1]{#1}%
\providecommand \enquote  [1]{``#1''}%
\providecommand \bibnamefont  [1]{#1}%
\providecommand \bibfnamefont [1]{#1}%
\providecommand \citenamefont [1]{#1}%
\providecommand \href@noop [0]{\@secondoftwo}%
\providecommand \href [0]{\begingroup \@sanitize@url \@href}%
\providecommand \@href[1]{\@@startlink{#1}\@@href}%
\providecommand \@@href[1]{\endgroup#1\@@endlink}%
\providecommand \@sanitize@url [0]{\catcode `\\12\catcode `\$12\catcode
  `\&12\catcode `\#12\catcode `\^12\catcode `\_12\catcode `\%12\relax}%
\providecommand \@@startlink[1]{}%
\providecommand \@@endlink[0]{}%
\providecommand \url  [0]{\begingroup\@sanitize@url \@url }%
\providecommand \@url [1]{\endgroup\@href {#1}{\urlprefix }}%
\providecommand \urlprefix  [0]{URL }%
\providecommand \Eprint [0]{\href }%
\providecommand \doibase [0]{http://dx.doi.org/}%
\providecommand \selectlanguage [0]{\@gobble}%
\providecommand \bibinfo  [0]{\@secondoftwo}%
\providecommand \bibfield  [0]{\@secondoftwo}%
\providecommand \translation [1]{[#1]}%
\providecommand \BibitemOpen [0]{}%
\providecommand \bibitemStop [0]{}%
\providecommand \bibitemNoStop [0]{.\EOS\space}%
\providecommand \EOS [0]{\spacefactor3000\relax}%
\providecommand \BibitemShut  [1]{\csname bibitem#1\endcsname}%
\let\auto@bib@innerbib\@empty
\bibitem [{\citenamefont {Unruh}(1976)}]{unruh_notes_1976}%
  \BibitemOpen
  \bibfield  {author} {\bibinfo {author} {\bibfnamefont {W.~G.}\ \bibnamefont
  {Unruh}},\ }\href {\doibase 10.1103/PhysRevD.14.870} {\bibfield  {journal}
  {\bibinfo  {journal} {Phys. Rev. D}\ }\textbf {\bibinfo {volume} {14}},\
  \bibinfo {pages} {870} (\bibinfo {year} {1976})}\BibitemShut {NoStop}%
\bibitem [{\citenamefont {DeWitt}(1979)}]{DeWitts}%
  \BibitemOpen
  \bibfield  {author} {\bibinfo {author} {\bibfnamefont {B.}~\bibnamefont
  {DeWitt}},\ }in\ \href@noop {} {\emph {\bibinfo {booktitle} {General
  Relativity: An Einstein Centenary Survey}}},\ \bibinfo {editor} {edited by\
  \bibinfo {editor} {\bibfnamefont {S.~W.}\ \bibnamefont {Hawking}}\ and\
  \bibinfo {editor} {\bibfnamefont {W.}~\bibnamefont {Israel}}}\ (\bibinfo
  {publisher} {Cambridge University Press},\ \bibinfo {address} {Cambridge},\
  \bibinfo {year} {1979})\BibitemShut {NoStop}%
\bibitem [{\citenamefont {Glauber}(1959)}]{Glauber}%
  \BibitemOpen
  \bibfield  {author} {\bibinfo {author} {\bibfnamefont {R.~J.}\ \bibnamefont
  {Glauber}},\ }\href@noop {} {\emph {\bibinfo {title} {Lectures in Theoretical
  Physics}}},\ edited by\ \bibinfo {editor} {\bibfnamefont {W.}~\bibnamefont
  {Brittin}}\ and\ \bibinfo {editor} {\bibfnamefont {L.}~\bibnamefont
  {Dunham}},\ Vol.\ \bibinfo {volume} {1:315}\ (\bibinfo  {publisher} {New
  York: Interscience},\ \bibinfo {year} {1959})\BibitemShut {NoStop}%
\bibitem [{\citenamefont {Jaynes}\ and\ \citenamefont
  {Cummings}(1963)}]{JaynesCumm}%
  \BibitemOpen
  \bibfield  {author} {\bibinfo {author} {\bibfnamefont {E.~T.}\ \bibnamefont
  {Jaynes}}\ and\ \bibinfo {author} {\bibfnamefont {F.~W.}\ \bibnamefont
  {Cummings}},\ }\href {\doibase 10.1109/PROC.1963.1664} {\bibfield  {journal}
  {\bibinfo  {journal} {Proceedings of the IEEE}\ }\textbf {\bibinfo {volume}
  {51}},\ \bibinfo {pages} {89} (\bibinfo {year} {1963})}\BibitemShut {NoStop}%
\bibitem [{\citenamefont {Scully}\ and\ \citenamefont
  {Zubairy}(1997)}]{scullybook}%
  \BibitemOpen
  \bibfield  {author} {\bibinfo {author} {\bibfnamefont {M.~O.}\ \bibnamefont
  {Scully}}\ and\ \bibinfo {author} {\bibfnamefont {M.~S.}\ \bibnamefont
  {Zubairy}},\ }\href@noop {} {\emph {\bibinfo {title} {Quantum Optics}}}\
  (\bibinfo  {publisher} {Cambridge University Press},\ \bibinfo {year}
  {1997})\BibitemShut {NoStop}%
\bibitem [{\citenamefont {Lin}(2012)}]{Lin2}%
  \BibitemOpen
  \bibfield  {author} {\bibinfo {author} {\bibfnamefont {S.-Y.}\ \bibnamefont
  {Lin}},\ }\href {\doibase https://doi.org/10.1016/j.aop.2012.08.004}
  {\bibfield  {journal} {\bibinfo  {journal} {Ann. Phys.}\ }\textbf {\bibinfo
  {volume} {327}},\ \bibinfo {pages} {3102 } (\bibinfo {year}
  {2012})}\BibitemShut {NoStop}%
\bibitem [{\citenamefont {Dragan}\ \emph
  {et~al.}(2013{\natexlab{a}})\citenamefont {Dragan}, \citenamefont {Doukas},\
  and\ \citenamefont {Mart\'{\i}n-Mart\'{\i}nez}}]{Drago1}%
  \BibitemOpen
  \bibfield  {author} {\bibinfo {author} {\bibfnamefont {A.}~\bibnamefont
  {Dragan}}, \bibinfo {author} {\bibfnamefont {J.}~\bibnamefont {Doukas}}, \
  and\ \bibinfo {author} {\bibfnamefont {E.}~\bibnamefont
  {Mart\'{\i}n-Mart\'{\i}nez}},\ }\href {\doibase 10.1103/PhysRevA.87.052326}
  {\bibfield  {journal} {\bibinfo  {journal} {Phys. Rev. A}\ }\textbf {\bibinfo
  {volume} {87}},\ \bibinfo {pages} {052326} (\bibinfo {year}
  {2013}{\natexlab{a}})}\BibitemShut {NoStop}%
\bibitem [{\citenamefont {Dragan}\ \emph
  {et~al.}(2013{\natexlab{b}})\citenamefont {Dragan}, \citenamefont {Doukas},
  \citenamefont {Mart{\'\i}n-Mart{\'\i}nez},\ and\ \citenamefont
  {Bruschi}}]{Drago2}%
  \BibitemOpen
  \bibfield  {author} {\bibinfo {author} {\bibfnamefont {A.}~\bibnamefont
  {Dragan}}, \bibinfo {author} {\bibfnamefont {J.}~\bibnamefont {Doukas}},
  \bibinfo {author} {\bibfnamefont {E.}~\bibnamefont
  {Mart{\'\i}n-Mart{\'\i}nez}}, \ and\ \bibinfo {author} {\bibfnamefont
  {D.~E.}\ \bibnamefont {Bruschi}},\ }\href
  {http://stacks.iop.org/0264-9381/30/i=23/a=235006} {\bibfield  {journal}
  {\bibinfo  {journal} {Classical and Quantum Gravity}\ }\textbf {\bibinfo
  {volume} {30}},\ \bibinfo {pages} {235006} (\bibinfo {year}
  {2013}{\natexlab{b}})}\BibitemShut {NoStop}%
\bibitem [{\citenamefont {Lin}(2014)}]{Lin2014773}%
  \BibitemOpen
  \bibfield  {author} {\bibinfo {author} {\bibfnamefont {S.-Y.}\ \bibnamefont
  {Lin}},\ }\href {\doibase 10.1016/j.aop.2014.08.018} {\bibfield  {journal}
  {\bibinfo  {journal} {Ann. Phys.}\ }\textbf {\bibinfo {volume} {351}},\
  \bibinfo {pages} {773 } (\bibinfo {year} {2014})}\BibitemShut {NoStop}%
\bibitem [{\citenamefont {Crispino}\ \emph {et~al.}(2008)\citenamefont
  {Crispino}, \citenamefont {Higuchi},\ and\ \citenamefont
  {Matsas}}]{Crispino}%
  \BibitemOpen
  \bibfield  {author} {\bibinfo {author} {\bibfnamefont {L.~C.~B.}\
  \bibnamefont {Crispino}}, \bibinfo {author} {\bibfnamefont {A.}~\bibnamefont
  {Higuchi}}, \ and\ \bibinfo {author} {\bibfnamefont {G.~E.~A.}\ \bibnamefont
  {Matsas}},\ }\href@noop {} {\bibfield  {journal} {\bibinfo  {journal} {Rev.
  Mod. Phys.}\ }\textbf {\bibinfo {volume} {80}},\ \bibinfo {pages} {787}
  (\bibinfo {year} {2008})}\BibitemShut {NoStop}%
\bibitem [{\citenamefont {Takagi}(1986)}]{Takagi}%
  \BibitemOpen
  \bibfield  {author} {\bibinfo {author} {\bibfnamefont {S.}~\bibnamefont
  {Takagi}},\ }\href@noop {} {\bibfield  {journal} {\bibinfo  {journal} {Prog.
  Theor. Phys. Suppl.}\ }\textbf {\bibinfo {volume} {88}},\ \bibinfo {pages}
  {1} (\bibinfo {year} {1986})}\BibitemShut {NoStop}%
\bibitem [{\citenamefont {Candelas}\ and\ \citenamefont
  {Sciama}(1977)}]{candelas_irreversible_1977}%
  \BibitemOpen
  \bibfield  {author} {\bibinfo {author} {\bibfnamefont {P.}~\bibnamefont
  {Candelas}}\ and\ \bibinfo {author} {\bibfnamefont {D.~W.}\ \bibnamefont
  {Sciama}},\ }\href {\doibase 10.1103/PhysRevLett.38.1372} {\bibfield
  {journal} {\bibinfo  {journal} {Phys. Rev. Lett.}\ }\textbf {\bibinfo
  {volume} {38}},\ \bibinfo {pages} {1372} (\bibinfo {year}
  {1977})}\BibitemShut {NoStop}%
\bibitem [{\citenamefont {Wallraff}\ \emph {et~al.}(2004)\citenamefont
  {Wallraff}, \citenamefont {Schuster}, \citenamefont {Blais}, \citenamefont
  {Frunzio}, \citenamefont {Huang}, \citenamefont {Majer}, \citenamefont
  {Kumar}, \citenamefont {Girvin},\ and\ \citenamefont
  {Schoelkopf}}]{Wallraff:2004aa}%
  \BibitemOpen
  \bibfield  {author} {\bibinfo {author} {\bibfnamefont {A.}~\bibnamefont
  {Wallraff}}, \bibinfo {author} {\bibfnamefont {D.~I.}\ \bibnamefont
  {Schuster}}, \bibinfo {author} {\bibfnamefont {A.}~\bibnamefont {Blais}},
  \bibinfo {author} {\bibfnamefont {L.}~\bibnamefont {Frunzio}}, \bibinfo
  {author} {\bibfnamefont {R.~S.}\ \bibnamefont {Huang}}, \bibinfo {author}
  {\bibfnamefont {J.}~\bibnamefont {Majer}}, \bibinfo {author} {\bibfnamefont
  {S.}~\bibnamefont {Kumar}}, \bibinfo {author} {\bibfnamefont {S.~M.}\
  \bibnamefont {Girvin}}, \ and\ \bibinfo {author} {\bibfnamefont {R.~J.}\
  \bibnamefont {Schoelkopf}},\ }\href {http://dx.doi.org/10.1038/nature02851}
  {\bibfield  {journal} {\bibinfo  {journal} {Nature}\ }\textbf {\bibinfo
  {volume} {431}},\ \bibinfo {pages} {162} (\bibinfo {year}
  {2004})}\BibitemShut {NoStop}%
\bibitem [{\citenamefont
  {Mart{\'{i}}n-Mart{\'{i}}nez}(2015)}]{Martin-Martinez2015}%
  \BibitemOpen
  \bibfield  {author} {\bibinfo {author} {\bibfnamefont {E.}~\bibnamefont
  {Mart{\'{i}}n-Mart{\'{i}}nez}},\ }\href {\doibase 10.1103/PhysRevD.92.104019}
  {\bibfield  {journal} {\bibinfo  {journal} {Phys. Rev. D}\ }\textbf {\bibinfo
  {volume} {92}},\ \bibinfo {pages} {104019} (\bibinfo {year}
  {2015})}\BibitemShut {NoStop}%
\bibitem [{\citenamefont {Pozas-Kerstjens}\ and\ \citenamefont
  {Mart\'{\i}n-Mart\'{\i}nez}(2016)}]{Pozas-Kerstjens2016}%
  \BibitemOpen
  \bibfield  {author} {\bibinfo {author} {\bibfnamefont {A.}~\bibnamefont
  {Pozas-Kerstjens}}\ and\ \bibinfo {author} {\bibfnamefont {E.}~\bibnamefont
  {Mart\'{\i}n-Mart\'{\i}nez}},\ }\href {\doibase 10.1103/PhysRevD.94.064074}
  {\bibfield  {journal} {\bibinfo  {journal} {Phys. Rev. D}\ }\textbf {\bibinfo
  {volume} {94}},\ \bibinfo {pages} {064074} (\bibinfo {year}
  {2016})}\BibitemShut {NoStop}%
\bibitem [{\citenamefont {Lopp}\ and\ \citenamefont
  {Mart\'in-Mart\'inez}(2018)}]{Richard}%
  \BibitemOpen
  \bibfield  {author} {\bibinfo {author} {\bibfnamefont {R.}~\bibnamefont
  {Lopp}}\ and\ \bibinfo {author} {\bibfnamefont {E.}~\bibnamefont
  {Mart\'in-Mart\'inez}},\ }\href {\doibase
  https://doi.org/10.1016/j.optcom.2018.03.056} {\bibfield  {journal} {\bibinfo
   {journal} {Opt. Commun.}\ }\textbf {\bibinfo {volume} {423}},\ \bibinfo
  {pages} {29 } (\bibinfo {year} {2018})}\BibitemShut {NoStop}%
\bibitem [{\citenamefont {Mart\'{i}n-Mart\'{i}nez}\ \emph
  {et~al.}(2013{\natexlab{a}})\citenamefont {Mart\'{i}n-Mart\'{i}nez},
  \citenamefont {Aasen},\ and\ \citenamefont {Kempf}}]{AasenPRL}%
  \BibitemOpen
  \bibfield  {author} {\bibinfo {author} {\bibfnamefont {E.}~\bibnamefont
  {Mart\'{i}n-Mart\'{i}nez}}, \bibinfo {author} {\bibfnamefont
  {D.}~\bibnamefont {Aasen}}, \ and\ \bibinfo {author} {\bibfnamefont
  {A.}~\bibnamefont {Kempf}},\ }\href {\doibase 10.1103/PhysRevLett.110.160501}
  {\bibfield  {journal} {\bibinfo  {journal} {Phys. Rev. Lett.}\ }\textbf
  {\bibinfo {volume} {110}},\ \bibinfo {pages} {160501} (\bibinfo {year}
  {2013}{\natexlab{a}})}\BibitemShut {NoStop}%
\bibitem [{\citenamefont {Lee}\ and\ \citenamefont {Fuentes}(2014)}]{Lee}%
  \BibitemOpen
  \bibfield  {author} {\bibinfo {author} {\bibfnamefont {A.~R.}\ \bibnamefont
  {Lee}}\ and\ \bibinfo {author} {\bibfnamefont {I.}~\bibnamefont {Fuentes}},\
  }\href {\doibase 10.1103/PhysRevD.89.085041} {\bibfield  {journal} {\bibinfo
  {journal} {Phys. Rev. D}\ }\textbf {\bibinfo {volume} {89}},\ \bibinfo
  {pages} {085041} (\bibinfo {year} {2014})}\BibitemShut {NoStop}%
\bibitem [{\citenamefont {Mart\'{i}n-Mart\'{i}nez}\ and\ \citenamefont
  {Sutherland}(2014)}]{Chris}%
  \BibitemOpen
  \bibfield  {author} {\bibinfo {author} {\bibfnamefont {E.}~\bibnamefont
  {Mart\'{i}n-Mart\'{i}nez}}\ and\ \bibinfo {author} {\bibfnamefont
  {C.}~\bibnamefont {Sutherland}},\ }\href {\doibase
  http://dx.doi.org/10.1016/j.physletb.2014.10.038} {\bibfield  {journal}
  {\bibinfo  {journal} {Phys. Lett. B}\ }\textbf {\bibinfo {volume} {739}},\
  \bibinfo {pages} {74 } (\bibinfo {year} {2014})}\BibitemShut {NoStop}%
\bibitem [{\citenamefont {Cliche}\ and\ \citenamefont
  {Kempf}(2010)}]{mathieuachim1}%
  \BibitemOpen
  \bibfield  {author} {\bibinfo {author} {\bibfnamefont {M.}~\bibnamefont
  {Cliche}}\ and\ \bibinfo {author} {\bibfnamefont {A.}~\bibnamefont {Kempf}},\
  }\href {\doibase 10.1103/PhysRevA.81.012330} {\bibfield  {journal} {\bibinfo
  {journal} {Phys. Rev. A}\ }\textbf {\bibinfo {volume} {81}},\ \bibinfo
  {pages} {012330} (\bibinfo {year} {2010})}\BibitemShut {NoStop}%
\bibitem [{\citenamefont {Jonsson}\ \emph {et~al.}(2015)\citenamefont
  {Jonsson}, \citenamefont {Mart\'{i}n-Mart\'{i}nez},\ and\ \citenamefont
  {Kempf}}]{Robort2}%
  \BibitemOpen
  \bibfield  {author} {\bibinfo {author} {\bibfnamefont {R.~H.}\ \bibnamefont
  {Jonsson}}, \bibinfo {author} {\bibfnamefont {E.}~\bibnamefont
  {Mart\'{i}n-Mart\'{i}nez}}, \ and\ \bibinfo {author} {\bibfnamefont
  {A.}~\bibnamefont {Kempf}},\ }\href {\doibase 10.1103/PhysRevLett.114.110505}
  {\bibfield  {journal} {\bibinfo  {journal} {Phys. Rev. Lett.}\ }\textbf
  {\bibinfo {volume} {114}},\ \bibinfo {pages} {110505} (\bibinfo {year}
  {2015})}\BibitemShut {NoStop}%
\bibitem [{\citenamefont {Landulfo}(2016)}]{Landulfo}%
  \BibitemOpen
  \bibfield  {author} {\bibinfo {author} {\bibfnamefont {A.~G.~S.}\
  \bibnamefont {Landulfo}},\ }\href {\doibase 10.1103/PhysRevD.93.104019}
  {\bibfield  {journal} {\bibinfo  {journal} {Phys. Rev. D}\ }\textbf {\bibinfo
  {volume} {93}},\ \bibinfo {pages} {104019} (\bibinfo {year}
  {2016})}\BibitemShut {NoStop}%
\bibitem [{\citenamefont {Jonsson}(2017)}]{Robort3}%
  \BibitemOpen
  \bibfield  {author} {\bibinfo {author} {\bibfnamefont {R.~H.}\ \bibnamefont
  {Jonsson}},\ }\href {http://stacks.iop.org/1751-8121/50/i=35/a=355401}
  {\bibfield  {journal} {\bibinfo  {journal} {J. Phys. A: Math. Theor.}\
  }\textbf {\bibinfo {volume} {50}},\ \bibinfo {pages} {355401} (\bibinfo
  {year} {2017})}\BibitemShut {NoStop}%
\bibitem [{\citenamefont {Gibbons}\ and\ \citenamefont
  {Hawking}(1977)}]{Gibbons1977}%
  \BibitemOpen
  \bibfield  {author} {\bibinfo {author} {\bibfnamefont {G.~W.}\ \bibnamefont
  {Gibbons}}\ and\ \bibinfo {author} {\bibfnamefont {S.~W.}\ \bibnamefont
  {Hawking}},\ }\href {\doibase 10.1103/PhysRevD.15.2738} {\bibfield  {journal}
  {\bibinfo  {journal} {Phys. Rev. D}\ }\textbf {\bibinfo {volume} {15}},\
  \bibinfo {pages} {2738} (\bibinfo {year} {1977})}\BibitemShut {NoStop}%
\bibitem [{\citenamefont {Mart\'{i}n-Mart\'{i}nez}\ and\ \citenamefont
  {Menicucci}(2012)}]{cosmoq}%
  \BibitemOpen
  \bibfield  {author} {\bibinfo {author} {\bibfnamefont {E.}~\bibnamefont
  {Mart\'{i}n-Mart\'{i}nez}}\ and\ \bibinfo {author} {\bibfnamefont {N.~C.}\
  \bibnamefont {Menicucci}},\ }\href {\doibase 10.1088/0264-9381/29/22/224003}
  {\bibfield  {journal} {\bibinfo  {journal} {Classical and Quantum Gravity}\
  }\textbf {\bibinfo {volume} {29}},\ \bibinfo {pages} {224003} (\bibinfo
  {year} {2012})}\BibitemShut {NoStop}%
\bibitem [{\citenamefont {Fukuma}\ \emph {et~al.}(2014)\citenamefont {Fukuma},
  \citenamefont {Sugishita},\ and\ \citenamefont {Sakatani}}]{FilCos}%
  \BibitemOpen
  \bibfield  {author} {\bibinfo {author} {\bibfnamefont {M.}~\bibnamefont
  {Fukuma}}, \bibinfo {author} {\bibfnamefont {S.}~\bibnamefont {Sugishita}}, \
  and\ \bibinfo {author} {\bibfnamefont {Y.}~\bibnamefont {Sakatani}},\ }\href
  {\doibase 10.1103/PhysRevD.89.064024} {\bibfield  {journal} {\bibinfo
  {journal} {Phys. Rev. D}\ }\textbf {\bibinfo {volume} {89}},\ \bibinfo
  {pages} {064024} (\bibinfo {year} {2014})}\BibitemShut {NoStop}%
\bibitem [{\citenamefont {Garay}\ \emph {et~al.}(2014)\citenamefont {Garay},
  \citenamefont {Mart\'in-Benito},\ and\ \citenamefont
  {Mart\'in-Mart\'inez}}]{QuanG}%
  \BibitemOpen
  \bibfield  {author} {\bibinfo {author} {\bibfnamefont {L.~J.}\ \bibnamefont
  {Garay}}, \bibinfo {author} {\bibfnamefont {M.}~\bibnamefont
  {Mart\'in-Benito}}, \ and\ \bibinfo {author} {\bibfnamefont {E.}~\bibnamefont
  {Mart\'in-Mart\'inez}},\ }\href {\doibase 10.1103/PhysRevD.89.043510}
  {\bibfield  {journal} {\bibinfo  {journal} {Phys. Rev. D}\ }\textbf {\bibinfo
  {volume} {89}},\ \bibinfo {pages} {043510} (\bibinfo {year}
  {2014})}\BibitemShut {NoStop}%
\bibitem [{\citenamefont {Blasco}\ \emph {et~al.}(2015)\citenamefont {Blasco},
  \citenamefont {Garay}, \citenamefont {Mart\'{\i}n-Benito},\ and\
  \citenamefont {Mart\'{\i}n-Mart\'{\i}nez}}]{Blasco}%
  \BibitemOpen
  \bibfield  {author} {\bibinfo {author} {\bibfnamefont {A.}~\bibnamefont
  {Blasco}}, \bibinfo {author} {\bibfnamefont {L.~J.}\ \bibnamefont {Garay}},
  \bibinfo {author} {\bibfnamefont {M.}~\bibnamefont {Mart\'{\i}n-Benito}}, \
  and\ \bibinfo {author} {\bibfnamefont {E.}~\bibnamefont
  {Mart\'{\i}n-Mart\'{\i}nez}},\ }\href {\doibase
  10.1103/PhysRevLett.114.141103} {\bibfield  {journal} {\bibinfo  {journal}
  {Phys. Rev. Lett.}\ }\textbf {\bibinfo {volume} {114}},\ \bibinfo {pages}
  {141103} (\bibinfo {year} {2015})}\BibitemShut {NoStop}%
\bibitem [{\citenamefont {Alhambra}\ \emph {et~al.}(2014)\citenamefont
  {Alhambra}, \citenamefont {Kempf},\ and\ \citenamefont
  {Mart{\'{i}}n-Mart{\'{i}}nez}}]{Alhambra2014}%
  \BibitemOpen
  \bibfield  {author} {\bibinfo {author} {\bibfnamefont {{\'{A}}.~M.}\
  \bibnamefont {Alhambra}}, \bibinfo {author} {\bibfnamefont {A.}~\bibnamefont
  {Kempf}}, \ and\ \bibinfo {author} {\bibfnamefont {E.}~\bibnamefont
  {Mart{\'{i}}n-Mart{\'{i}}nez}},\ }\href {\doibase 10.1103/PhysRevA.89.033835}
  {\bibfield  {journal} {\bibinfo  {journal} {Phys. Rev. A}\ }\textbf {\bibinfo
  {volume} {89}},\ \bibinfo {pages} {033835} (\bibinfo {year}
  {2014})}\BibitemShut {NoStop}%
\bibitem [{\citenamefont {Marino}\ \emph {et~al.}(2014)\citenamefont {Marino},
  \citenamefont {Noto},\ and\ \citenamefont {Passante}}]{Marino2014}%
  \BibitemOpen
  \bibfield  {author} {\bibinfo {author} {\bibfnamefont {J.}~\bibnamefont
  {Marino}}, \bibinfo {author} {\bibfnamefont {A.}~\bibnamefont {Noto}}, \ and\
  \bibinfo {author} {\bibfnamefont {R.}~\bibnamefont {Passante}},\ }\href
  {\doibase 10.1103/PhysRevLett.113.020403} {\bibfield  {journal} {\bibinfo
  {journal} {Phys. Rev. Lett.}\ }\textbf {\bibinfo {volume} {113}},\ \bibinfo
  {pages} {020403} (\bibinfo {year} {2014})}\BibitemShut {NoStop}%
\bibitem [{\citenamefont {Intravaia}\ \emph {et~al.}(2015)\citenamefont
  {Intravaia}, \citenamefont {Mkrtchian}, \citenamefont {Buhmann},
  \citenamefont {Scheel}, \citenamefont {Dalvit},\ and\ \citenamefont
  {Henkel}}]{Dalvitian}%
  \BibitemOpen
  \bibfield  {author} {\bibinfo {author} {\bibfnamefont {F.}~\bibnamefont
  {Intravaia}}, \bibinfo {author} {\bibfnamefont {V.~E.}\ \bibnamefont
  {Mkrtchian}}, \bibinfo {author} {\bibfnamefont {S.~Y.}\ \bibnamefont
  {Buhmann}}, \bibinfo {author} {\bibfnamefont {S.}~\bibnamefont {Scheel}},
  \bibinfo {author} {\bibfnamefont {D.~A.~R.}\ \bibnamefont {Dalvit}}, \ and\
  \bibinfo {author} {\bibfnamefont {C.}~\bibnamefont {Henkel}},\ }\href
  {http://stacks.iop.org/0953-8984/27/i=21/a=214020} {\bibfield  {journal}
  {\bibinfo  {journal} {Journal of Physics: Condensed Matter}\ }\textbf
  {\bibinfo {volume} {27}},\ \bibinfo {pages} {214020} (\bibinfo {year}
  {2015})}\BibitemShut {NoStop}%
\bibitem [{\citenamefont {Valentini}(1991)}]{Valentini1991}%
  \BibitemOpen
  \bibfield  {author} {\bibinfo {author} {\bibfnamefont {A.}~\bibnamefont
  {Valentini}},\ }\href {\doibase
  http://dx.doi.org/10.1016/0375-9601(91)90952-5} {\bibfield  {journal}
  {\bibinfo  {journal} {Physics Letters A}\ }\textbf {\bibinfo {volume}
  {153}},\ \bibinfo {pages} {321 } (\bibinfo {year} {1991})}\BibitemShut
  {NoStop}%
\bibitem [{\citenamefont {Reznik}(2003)}]{Reznik2003}%
  \BibitemOpen
  \bibfield  {author} {\bibinfo {author} {\bibfnamefont {B.}~\bibnamefont
  {Reznik}},\ }\href {\doibase 10.1023/A:1022875910744} {\bibfield  {journal}
  {\bibinfo  {journal} {Foundations of Physics}\ }\textbf {\bibinfo {volume}
  {33}},\ \bibinfo {pages} {167} (\bibinfo {year} {2003})}\BibitemShut
  {NoStop}%
\bibitem [{\citenamefont {Reznik}\ \emph {et~al.}(2005)\citenamefont {Reznik},
  \citenamefont {Retzker},\ and\ \citenamefont {Silman}}]{Reznik1}%
  \BibitemOpen
  \bibfield  {author} {\bibinfo {author} {\bibfnamefont {B.}~\bibnamefont
  {Reznik}}, \bibinfo {author} {\bibfnamefont {A.}~\bibnamefont {Retzker}}, \
  and\ \bibinfo {author} {\bibfnamefont {J.}~\bibnamefont {Silman}},\ }\href
  {http://link.aps.org/abstract/PRA/v71/e042104} {\bibfield  {journal}
  {\bibinfo  {journal} {Phys. Rev. A}\ }\textbf {\bibinfo {volume} {71}},\
  \bibinfo {eid} {042104} (\bibinfo {year} {2005})}\BibitemShut {NoStop}%
\bibitem [{\citenamefont {Ver~Steeg}\ and\ \citenamefont
  {Menicucci}(2009)}]{VerSteeg2009}%
  \BibitemOpen
  \bibfield  {author} {\bibinfo {author} {\bibfnamefont {G.}~\bibnamefont
  {Ver~Steeg}}\ and\ \bibinfo {author} {\bibfnamefont {N.~C.}\ \bibnamefont
  {Menicucci}},\ }\href {\doibase 10.1103/PhysRevD.79.044027} {\bibfield
  {journal} {\bibinfo  {journal} {Phys. Rev. D}\ }\textbf {\bibinfo {volume}
  {79}},\ \bibinfo {pages} {044027} (\bibinfo {year} {2009})}\BibitemShut
  {NoStop}%
\bibitem [{\citenamefont {Olson}\ and\ \citenamefont
  {Ralph}(2011)}]{Olson2011}%
  \BibitemOpen
  \bibfield  {author} {\bibinfo {author} {\bibfnamefont {S.~J.}\ \bibnamefont
  {Olson}}\ and\ \bibinfo {author} {\bibfnamefont {T.~C.}\ \bibnamefont
  {Ralph}},\ }\href {\doibase 10.1103/PhysRevLett.106.110404} {\bibfield
  {journal} {\bibinfo  {journal} {Phys. Rev. Lett.}\ }\textbf {\bibinfo
  {volume} {106}},\ \bibinfo {pages} {110404} (\bibinfo {year}
  {2011})}\BibitemShut {NoStop}%
\bibitem [{\citenamefont {Olson}\ and\ \citenamefont {Ralph}(2012)}]{Oslon2}%
  \BibitemOpen
  \bibfield  {author} {\bibinfo {author} {\bibfnamefont {S.~J.}\ \bibnamefont
  {Olson}}\ and\ \bibinfo {author} {\bibfnamefont {T.~C.}\ \bibnamefont
  {Ralph}},\ }\href {\doibase 10.1103/PhysRevA.85.012306} {\bibfield  {journal}
  {\bibinfo  {journal} {Phys. Rev. A}\ }\textbf {\bibinfo {volume} {85}},\
  \bibinfo {pages} {012306} (\bibinfo {year} {2012})}\BibitemShut {NoStop}%
\bibitem [{\citenamefont {Salton}\ \emph {et~al.}(2015)\citenamefont {Salton},
  \citenamefont {Mann},\ and\ \citenamefont {Menicucci}}]{Salton:2014jaa}%
  \BibitemOpen
  \bibfield  {author} {\bibinfo {author} {\bibfnamefont {G.}~\bibnamefont
  {Salton}}, \bibinfo {author} {\bibfnamefont {R.~B.}\ \bibnamefont {Mann}}, \
  and\ \bibinfo {author} {\bibfnamefont {N.~C.}\ \bibnamefont {Menicucci}},\
  }\href {\doibase 10.1088/1367-2630/17/3/035001} {\bibfield  {journal}
  {\bibinfo  {journal} {New J. Phys.}\ }\textbf {\bibinfo {volume} {17}},\
  \bibinfo {pages} {035001} (\bibinfo {year} {2015})}\BibitemShut {NoStop}%
\bibitem [{\citenamefont {Pozas-Kerstjens}\ and\ \citenamefont
  {Mart\'{\i}n-Mart\'{\i}nez}(2015)}]{Pozas-Kerstjens:2015}%
  \BibitemOpen
  \bibfield  {author} {\bibinfo {author} {\bibfnamefont {A.}~\bibnamefont
  {Pozas-Kerstjens}}\ and\ \bibinfo {author} {\bibfnamefont {E.}~\bibnamefont
  {Mart\'{\i}n-Mart\'{\i}nez}},\ }\href {\doibase 10.1103/PhysRevD.92.064042}
  {\bibfield  {journal} {\bibinfo  {journal} {Phys. Rev. D}\ }\textbf {\bibinfo
  {volume} {92}},\ \bibinfo {pages} {064042} (\bibinfo {year}
  {2015})}\BibitemShut {NoStop}%
\bibitem [{\citenamefont {Kukita}\ and\ \citenamefont {Nambu}(2017)}]{Yasuada}%
  \BibitemOpen
  \bibfield  {author} {\bibinfo {author} {\bibfnamefont {S.}~\bibnamefont
  {Kukita}}\ and\ \bibinfo {author} {\bibfnamefont {Y.}~\bibnamefont {Nambu}},\
  }\href {http://www.mdpi.com/1099-4300/19/9/449} {\bibfield  {journal}
  {\bibinfo  {journal} {Entropy}\ }\textbf {\bibinfo {volume} {19}},\ \bibinfo
  {pages} {449} (\bibinfo {year} {2017})}\BibitemShut {NoStop}%
\bibitem [{\citenamefont {Mart\'{i}n-Mart\'{i}nez}\ \emph
  {et~al.}(2013{\natexlab{b}})\citenamefont {Mart\'{i}n-Mart\'{i}nez},
  \citenamefont {Brown}, \citenamefont {Donnelly},\ and\ \citenamefont
  {Kempf}}]{Farming}%
  \BibitemOpen
  \bibfield  {author} {\bibinfo {author} {\bibfnamefont {E.}~\bibnamefont
  {Mart\'{i}n-Mart\'{i}nez}}, \bibinfo {author} {\bibfnamefont {E.~G.}\
  \bibnamefont {Brown}}, \bibinfo {author} {\bibfnamefont {W.}~\bibnamefont
  {Donnelly}}, \ and\ \bibinfo {author} {\bibfnamefont {A.}~\bibnamefont
  {Kempf}},\ }\href {\doibase 10.1103/PhysRevA.88.052310} {\bibfield  {journal}
  {\bibinfo  {journal} {Phys. Rev. A}\ }\textbf {\bibinfo {volume} {88}},\
  \bibinfo {pages} {052310} (\bibinfo {year} {2013}{\natexlab{b}})}\BibitemShut
  {NoStop}%
\bibitem [{\citenamefont {Brown}\ \emph {et~al.}(2014)\citenamefont {Brown},
  \citenamefont {Donnelly}, \citenamefont {Kempf}, \citenamefont {Mann},
  \citenamefont {Mart{\'\i}n-Mart{\'\i}nez},\ and\ \citenamefont
  {Menicucci}}]{Brown:2014en}%
  \BibitemOpen
  \bibfield  {author} {\bibinfo {author} {\bibfnamefont {E.~G.}\ \bibnamefont
  {Brown}}, \bibinfo {author} {\bibfnamefont {W.}~\bibnamefont {Donnelly}},
  \bibinfo {author} {\bibfnamefont {A.}~\bibnamefont {Kempf}}, \bibinfo
  {author} {\bibfnamefont {R.~B.}\ \bibnamefont {Mann}}, \bibinfo {author}
  {\bibfnamefont {E.}~\bibnamefont {Mart{\'\i}n-Mart{\'\i}nez}}, \ and\
  \bibinfo {author} {\bibfnamefont {N.~C.}\ \bibnamefont {Menicucci}},\ }\href
  {\doibase 10.1088/1367-2630/16/10/105020} {\bibfield  {journal} {\bibinfo
  {journal} {New J. Phys.}\ }\textbf {\bibinfo {volume} {16}},\ \bibinfo
  {pages} {105020} (\bibinfo {year} {2014})}\BibitemShut {NoStop}%
\bibitem [{\citenamefont {Summers}\ and\ \citenamefont
  {Werner}(1985)}]{Alegbra1}%
  \BibitemOpen
  \bibfield  {author} {\bibinfo {author} {\bibfnamefont {S.~J.}\ \bibnamefont
  {Summers}}\ and\ \bibinfo {author} {\bibfnamefont {R.~F.}\ \bibnamefont
  {Werner}},\ }\href@noop {} {\bibfield  {journal} {\bibinfo  {journal} {Phys.
  Lett. A}\ }\textbf {\bibinfo {volume} {110}},\ \bibinfo {pages} {257}
  (\bibinfo {year} {1985})}\BibitemShut {NoStop}%
\bibitem [{\citenamefont {Summers}\ and\ \citenamefont
  {Werner}(1987)}]{Alegbra2}%
  \BibitemOpen
  \bibfield  {author} {\bibinfo {author} {\bibfnamefont {S.~J.}\ \bibnamefont
  {Summers}}\ and\ \bibinfo {author} {\bibfnamefont {R.~F.}\ \bibnamefont
  {Werner}},\ }\href@noop {} {\bibfield  {journal} {\bibinfo  {journal} {J.
  Math. Phys.}\ }\textbf {\bibinfo {volume} {28}},\ \bibinfo {pages} {2440}
  (\bibinfo {year} {1987})}\BibitemShut {NoStop}%
\bibitem [{\citenamefont {Louko}\ and\ \citenamefont {Satz}(2006)}]{Satz2006}%
  \BibitemOpen
  \bibfield  {author} {\bibinfo {author} {\bibfnamefont {J.}~\bibnamefont
  {Louko}}\ and\ \bibinfo {author} {\bibfnamefont {A.}~\bibnamefont {Satz}},\
  }\href@noop {} {\bibfield  {journal} {\bibinfo  {journal} {Class. Quant.
  Grav.}\ }\textbf {\bibinfo {volume} {23}},\ \bibinfo {pages} {6321} (\bibinfo
  {year} {2006})}\BibitemShut {NoStop}%
\bibitem [{\citenamefont {Satz}(2007)}]{Satz2007}%
  \BibitemOpen
  \bibfield  {author} {\bibinfo {author} {\bibfnamefont {A.}~\bibnamefont
  {Satz}},\ }\href {\doibase 10.1088/0264-9381/24/7/003} {\bibfield  {journal}
  {\bibinfo  {journal} {Class. Quant. Grav.}\ }\textbf {\bibinfo {volume}
  {24}},\ \bibinfo {pages} {1719} (\bibinfo {year} {2007})}\BibitemShut
  {NoStop}%
\bibitem [{\citenamefont {Hodgkinson}\ and\ \citenamefont
  {Louko}(2012)}]{Hodgkinsonclick}%
  \BibitemOpen
  \bibfield  {author} {\bibinfo {author} {\bibfnamefont {L.}~\bibnamefont
  {Hodgkinson}}\ and\ \bibinfo {author} {\bibfnamefont {J.}~\bibnamefont
  {Louko}},\ }\href@noop {} {\bibfield  {journal} {\bibinfo  {journal} {J.
  Math. Phys.}\ }\textbf {\bibinfo {volume} {63}},\ \bibinfo {pages} {082301}
  (\bibinfo {year} {2012})}\BibitemShut {NoStop}%
\bibitem [{\citenamefont {Iyer}\ and\ \citenamefont
  {Kumar}(1980)}]{Takahashi2011}%
  \BibitemOpen
  \bibfield  {author} {\bibinfo {author} {\bibfnamefont {B.~R.}\ \bibnamefont
  {Iyer}}\ and\ \bibinfo {author} {\bibfnamefont {A.}~\bibnamefont {Kumar}},\
  }\href {http://stacks.iop.org/0305-4470/13/i=2/a=015} {\bibfield  {journal}
  {\bibinfo  {journal} {J. Phys. A}\ }\textbf {\bibinfo {volume} {13}},\
  \bibinfo {pages} {469} (\bibinfo {year} {1980})}\BibitemShut {NoStop}%
\bibitem [{\citenamefont {H\"ummer}\ \emph {et~al.}(2016)\citenamefont
  {H\"ummer}, \citenamefont {Mart\'{\i}n-Mart\'{\i}nez},\ and\ \citenamefont
  {Kempf}}]{hummer}%
  \BibitemOpen
  \bibfield  {author} {\bibinfo {author} {\bibfnamefont {D.}~\bibnamefont
  {H\"ummer}}, \bibinfo {author} {\bibfnamefont {E.}~\bibnamefont
  {Mart\'{\i}n-Mart\'{\i}nez}}, \ and\ \bibinfo {author} {\bibfnamefont
  {A.}~\bibnamefont {Kempf}},\ }\href {\doibase 10.1103/PhysRevD.93.024019}
  {\bibfield  {journal} {\bibinfo  {journal} {Phys. Rev. D}\ }\textbf {\bibinfo
  {volume} {93}},\ \bibinfo {pages} {024019} (\bibinfo {year}
  {2016})}\BibitemShut {NoStop}%
\bibitem [{\citenamefont {Ballentine}(1998)}]{BallentineB}%
  \BibitemOpen
  \bibfield  {author} {\bibinfo {author} {\bibfnamefont {L.~E.}\ \bibnamefont
  {Ballentine}},\ }\enquote {\bibinfo {title} {Quantum mechanics: A modern
  development},}\ \ (\bibinfo  {publisher} {World Scientific},\ \bibinfo {year}
  {1998})\ Chap.~\bibinfo {chapter} {9}\BibitemShut {NoStop}%
\bibitem [{\citenamefont {Kazantsev}(1974)}]{ruso}%
  \BibitemOpen
  \bibfield  {author} {\bibinfo {author} {\bibfnamefont {A.~P.}\ \bibnamefont
  {Kazantsev}},\ }\href@noop {} {\bibfield  {journal} {\bibinfo  {journal} {Zh.
  Eksp. Teor. Fiz.}\ }\textbf {\bibinfo {volume} {66}},\ \bibinfo {pages}
  {1599} (\bibinfo {year} {1974})}\BibitemShut {NoStop}%
\bibitem [{\citenamefont {Schlicht}(2004)}]{Schlicht2004}%
  \BibitemOpen
  \bibfield  {author} {\bibinfo {author} {\bibfnamefont {S.}~\bibnamefont
  {Schlicht}},\ }\href {\doibase 10.1088/0264-9381/21/19/011} {\bibfield
  {journal} {\bibinfo  {journal} {Classical and Quantum Gravity}\ }\textbf
  {\bibinfo {volume} {21}},\ \bibinfo {pages} {4647} (\bibinfo {year}
  {2004})}\BibitemShut {NoStop}%
\bibitem [{\citenamefont {Peropadre}\ \emph {et~al.}(2010)\citenamefont
  {Peropadre}, \citenamefont {Forn-D\'{\i}az}, \citenamefont {Solano},\ and\
  \citenamefont {Garc\'{\i}a-Ripoll}}]{Peropadre2010}%
  \BibitemOpen
  \bibfield  {author} {\bibinfo {author} {\bibfnamefont {B.}~\bibnamefont
  {Peropadre}}, \bibinfo {author} {\bibfnamefont {P.}~\bibnamefont
  {Forn-D\'{\i}az}}, \bibinfo {author} {\bibfnamefont {E.}~\bibnamefont
  {Solano}}, \ and\ \bibinfo {author} {\bibfnamefont {J.~J.}\ \bibnamefont
  {Garc\'{\i}a-Ripoll}},\ }\href {\doibase 10.1103/PhysRevLett.105.023601}
  {\bibfield  {journal} {\bibinfo  {journal} {Phys. Rev. Lett.}\ }\textbf
  {\bibinfo {volume} {105}},\ \bibinfo {pages} {023601} (\bibinfo {year}
  {2010})}\BibitemShut {NoStop}%
\bibitem [{\citenamefont {Forn-Diaz}\ \emph {et~al.}(2017)\citenamefont
  {Forn-Diaz}, \citenamefont {Garcia-Ripoll}, \citenamefont {Peropadre},
  \citenamefont {Orgiazzi}, \citenamefont {Yurtalan}, \citenamefont
  {Belyansky}, \citenamefont {Wilson},\ and\ \citenamefont
  {Lupascu}}]{Fornada}%
  \BibitemOpen
  \bibfield  {author} {\bibinfo {author} {\bibfnamefont {P.}~\bibnamefont
  {Forn-Diaz}}, \bibinfo {author} {\bibfnamefont {J.~J.}\ \bibnamefont
  {Garcia-Ripoll}}, \bibinfo {author} {\bibfnamefont {B.}~\bibnamefont
  {Peropadre}}, \bibinfo {author} {\bibfnamefont {J.~L.}\ \bibnamefont
  {Orgiazzi}}, \bibinfo {author} {\bibfnamefont {M.~A.}\ \bibnamefont
  {Yurtalan}}, \bibinfo {author} {\bibfnamefont {R.}~\bibnamefont {Belyansky}},
  \bibinfo {author} {\bibfnamefont {C.~M.}\ \bibnamefont {Wilson}}, \ and\
  \bibinfo {author} {\bibfnamefont {A.}~\bibnamefont {Lupascu}},\ }\href
  {http://dx.doi.org/10.1038/nphys3905} {\bibfield  {journal} {\bibinfo
  {journal} {Nat Phys}\ }\textbf {\bibinfo {volume} {13}},\ \bibinfo {pages}
  {39} (\bibinfo {year} {2017})}\BibitemShut {NoStop}%
\end{thebibliography}%

\end{document}